\documentclass[12pt,a4paper]{article}

\usepackage{epsfig}
\usepackage{cite} 
\usepackage{enumerate}

\usepackage{bold-extra}

\newlength{\abstwidth}
\newlength{\captivewidth}
\begin{document}
 
\renewcommand{\topfraction}{0.9}    
\renewcommand{\bottomfraction}{0.9} 
\renewcommand{\textfraction}{0.1}   
\renewcommand{\floatpagefraction}{0.8}
 
\newcommand{\mrm}[1]{\mathrm{#1}}
\newcommand{\mbf}[1]{\mathbf{#1}}
\newcommand{\mtt}[1]{\mathtt{#1}}
\newcommand{\tsc}[1]{\textsc{#1}}
\newcommand{\tbf}[1]{\textbf{#1}}
\newcommand{\ttt}[1]{\texttt{#1}}
\newcommand{\br}[1]{\overline{#1}}
\newlength{\tmplen}
\newcommand{\clab}[1]{\tiny\settowidth{\tmplen}{\scriptsize#1}%
\colorbox{white}{\textcolor{white}{#1}}\hspace*{-1.27\tmplen}\scriptsize#1}

\def\lsim{\mathrel{\rlap{\lower4pt\hbox{\hskip1pt$\sim$}}
    \raise1pt\hbox{$<$}}}                
\def\gsim{\mathrel{\rlap{\lower4pt\hbox{\hskip1pt$\sim$}}
    \raise1pt\hbox{$>$}}}                
\newcommand{\half}{$\frac{1}{2}$}        
 
\newcommand{\alphas}{\alpha_{\mathrm{s}}}
\newcommand{\alphaem}{\alpha_{\mathrm{em}}}
\newcommand{\pT}{\ensuremath{p_{\perp}}}
\newcommand{\kT}{\ensuremath{k_{\perp}}}
\newcommand{\pTmin}{p_{\perp\mathrm{min}}}
\newcommand{\pTo}{p_{\perp 0}}
\newcommand{\pTevol}{p_{\perp\mathrm{evol}}}
\newcommand{\pTPOW}{p_{\perp\mathrm{POWHEG}}}
\newcommand{\ECM}{E_{\mathrm{CM}}}
\newcommand{\mmin}{\mathrm{min}}
\newcommand{\mmax}{\mathrm{max}}
\newcommand{\GeV}{\ensuremath{\!\ \mathrm{GeV}}}
\newcommand{\TeV}{\ensuremath{\!\ \mathrm{TeV}}}
\newcommand{\rem}{\ensuremath{\mathrm{rem}}}
 
\renewcommand{\b}{\mathrm{b}}
\renewcommand{\c}{\mathrm{c}}
\renewcommand{\d}{\mathrm{d}}
\newcommand{\e}{\mathrm{e}}
\newcommand{\f}{\mathrm{f}}
\newcommand{\g}{\mathrm{g}}
\renewcommand{\j}{\mathrm{j}}
\newcommand{\J}{\mathrm{J}}
\newcommand{\hrm}{\mathrm{h}}
\newcommand{\n}{\mathrm{n}}
\newcommand{\p}{\mathrm{p}}
\newcommand{\q}{\mathrm{q}}
\newcommand{\s}{\mathrm{s}}
\renewcommand{\t}{\mathrm{t}}
\renewcommand{\u}{\mathrm{u}}
\newcommand{\A}{\mathrm{A}}
\newcommand{\B}{\mathrm{B}}
\newcommand{\D}{\mathrm{D}}
\renewcommand{\H}{\mathrm{H}}
\newcommand{\K}{\mathrm{K}}
\newcommand{\Q}{\mathrm{Q}}
\newcommand{\W}{\mathrm{W}}
\newcommand{\Z}{\mathrm{Z}}
\newcommand{\bbar}{\overline{\mathrm{b}}}
\newcommand{\cbar}{\overline{\mathrm{c}}}
\newcommand{\dbar}{\overline{\mathrm{d}}}
\newcommand{\fbar}{\overline{\mathrm{f}}}
\newcommand{\nbar}{\overline{\mathrm{n}}}
\newcommand{\pbar}{\overline{\mathrm{p}}}
\newcommand{\qbar}{\overline{\mathrm{q}}}
\newcommand{\sbar}{\overline{\mathrm{s}}}
\newcommand{\tbar}{\overline{\mathrm{t}}}
\newcommand{\ubar}{\overline{\mathrm{u}}}
\newcommand{\Bbar}{\overline{\mathrm{B}}}
\newcommand{\Dbar}{\overline{\mathrm{D}}}
\newcommand{\Qbar}{\overline{\mathrm{Q}}}
\newcommand{\qval}{\ensuremath{\q_{\mrm{v}}}}
\newcommand{\qsea}{\ensuremath{\q_{\mrm{s}}}}
\newcommand{\qcmp}{\ensuremath{\q_{\mrm{c}}}}
\newcommand{\val}{\ensuremath{{\mrm{v}}}}
\newcommand{\sea}{\ensuremath{{\mrm{s}}}}
\newcommand{\cmp}{\ensuremath{{\mrm{c}}}}
 
\newcommand{\sg}{\tilde{\mathrm{g}}}
\newcommand{\sq}{\tilde{\mathrm{q}}}
\newcommand{\sqd}{\tilde{\mathrm{d}}}
\newcommand{\squ}{\tilde{\mathrm{u}}}
\newcommand{\sqc}{\tilde{\mathrm{c}}}
\newcommand{\sqs}{\tilde{\mathrm{s}}}
\newcommand{\st}{\tilde{\mathrm{t}}}
\newcommand{\schi}{\tilde{\chi}}
 
\newenvironment{Itemize}{\begin{list}{$\bullet$}%
{\setlength{\topsep}{0.2mm}\setlength{\partopsep}{0.2mm}%
\setlength{\itemsep}{0.2mm}\setlength{\parsep}{0.2mm}}}%
{\end{list}}
\newcounter{enumct}
\newenvironment{Enumerate}{\begin{list}{\arabic{enumct}.}%
{\usecounter{enumct}\setlength{\topsep}{0.2mm}%
\setlength{\partopsep}{0.2mm}\setlength{\itemsep}{0.2mm}%
\setlength{\parsep}{0.2mm}}}{\end{list}}
 
\sloppy
 
\pagestyle{empty}
 
\begin{flushright}
LU TP 10-07\\
MCnet/10/04\\
March 2010
\end{flushright}
 
\vspace{\fill}
 
\begin{center}
\renewcommand{\thefootnote}{\fnsymbol{footnote}} 
{\LARGE\bf Improved Parton Showers}\\[3mm]
{\LARGE\bf at Large Transverse Momenta}%
\footnote{Work supported by the Marie Curie Early Stage Training program
``HEP-EST'' (contract number \makebox{MEST-CT-2005-019626}) and in part by
the Marie Curie RTN ``MCnet'' (contract number
\makebox{MRTN-CT-2006-035606})}\\[10mm]
\renewcommand{\thefootnote}{\arabic{footnote}}
\setcounter{footnote}{0}
{\Large R.~Corke\footnote{richard.corke@thep.lu.se} and %
T.~Sj\"ostrand\footnote{torbjorn@thep.lu.se}} \\[3mm]
{\it Theoretical High Energy Physics,}\\[1mm]
{\it Department of Astronomy and Theoretical Physics,}\\[1mm]
{\it Lund University,}\\[1mm]
{\it S\"olvegatan 14A,}\\[1mm]
{\it S-223 62 Lund, Sweden}
\end{center}
 
\vspace{\fill}
 
\begin{center}
{\bf Abstract}\\[2ex]
\begin{minipage}{\abstwidth}
Several methods to improve the parton-shower description of hard 
processes by an injection of matrix-element-based information 
have been presented over the years. In this article we study
(re)weighting schemes for the first/hardest emission. One objective
is to provide a consistent matching of the POWHEG next-to-leading 
order generator to the \tsc{Pythia} shower algorithms. Another is to
correct the default behaviour of these showers at large transverse 
momenta, based on a comparison with real-emission matrix elements.
\end{minipage}
\end{center}
 
\vspace{\fill}
 
\clearpage
\pagestyle{plain}
\setcounter{page}{1}

\section{Introduction}
With the start of the LHC in mind, there has been a recent focus on 
improving the description of event topologies and cross sections,
going beyond the Born level for many processes of interest.
Firstly, it involves the effects of events with one or more 
extra jets in the final state, which may affect the impact of 
background processes and thereby the choice of analysis strategies.
Secondly, it includes next-to-leading order (NLO) 
corrections to production cross sections, which are needed for
precision tests of the Standard Model and, hopefully, of physics 
beyond it.  

For the first point, real-emission matrix element (ME) calculations 
give a good description of hard and widely separated jets, while 
parton shower (PS) models give the correct behaviour in the soft and 
collinear regions of phase space. The goal is to find a way 
to combine these two methods, so that each is used in its region of 
validity, with a smooth transition between the two in all physical
distributions. This is not a trivial task. One key issue is that ME 
calculations describe inclusive events, while the PS generates
exclusive ones. The CKKW \cite{Catani:2001cc} method solves this
by using ME's supplemented by analytical Sudakov form factors to
go from an inclusive to an exclusive language in the hard region,
and then switching to a PS below some ME cutoff scale. In CKKW-L 
\cite{Lonnblad:2001iq} the Sudakovs are instead generated from 
fictitious showers using the same PS algorithm as in the soft region, 
to improve the consistency and continuity. These methods continue to evolve
\cite{Hoeche:2009rj,Hamilton:2009ne} while other approaches
include MLM \cite{MLM} and Pseudo-showers \cite{Mrenna:2003if}. Comparisons
between the methods have been made \cite{Alwall:2007fs,Lavesson:2007uu}.

For the second point, the virtual correction terms required at NLO
make the calculations more complicated. Analytically 
the cancellation of real and virtual ME divergences occur in the 
soft/collinear region, i.e.\ where we would rather want to use the PS 
description. The first approach to solve this issue for nontrivial
cases was MC@NLO \cite{Frixione:2002ik, Frixione:2003ei}. Here the
analytical expression is derived for the phase space population  
by the first shower emission, in the absence of a Sudakov form-factor
correction. The difference between the real-emission ME and this
analytical PS expression, which should be finite in the soft and
collinear limits, defines the differential cross section for events
with one real ``ME-based'' emission, from which the shower should start.
The rest of the cross section, wherein the analytical PS divergences 
and the virtual divergences cancel, gives the events where the shower 
should start from the lowest-order process. 

The MC@NLO approach has the disadvantage that the PS emission rate 
may well be higher than the ME one in some parts of phase space, 
in which case one is forced to introduce negative-weight events.
Furthermore, the analytical shower expressions are specific to a 
particular PS algorithm, so lengthy work has to be redone not only
between different generators but also for minor changes inside a
given generator \cite{LatundeDada:2007jg}. Both of these issues are 
solved with the POWHEG method \cite{Nason:2004rx,Frixione:2007vw}, where
the ME's themselves are exponentiated to provide a process-dependent
Sudakov form factor. Thereby, a positive-weight algorithm can be obtained
wherein POWHEG (almost) always generates one emission, chosen to be the one
with largest transverse momentum. It is then up to the subsequent shower to
respect this constraint, but otherwise without the need for a tight
connection between the ME and PS stages.

When POWHEG is used with showers that are not $\pT$-ordered, then
further thought must be given to the interface. Specifically, in  HERWIG, 
with its angular ordered showers \cite{Marchesini:1983bm, 
Marchesini:1987cf}, the first emission is at largest angle but not 
necessarily at largest $\pT$. This leads to the idea of a ``truncated 
shower'' \cite{Nason:2004rx}, where the shower is
modified such that the hardest emission is generated with a modified
Sudakov form factor. Subsequent emissions are then generated with the usual
algorithm, but with a $\pT$ veto so that they cannot be harder than the
hardest emission.

The POWHEG approach is especially convenient, then, if the shower itself is 
$\pT$-ordered, as is now the case in the \tsc{Pythia} generator 
\cite{Sjostrand:2004ef, Sjostrand:2006za, Sjostrand:2007gs}. 
Nevertheless there are some subtleties that should be taken into account to
optimise the interface. We discuss these issues in Section 2, and in
particular interface to POWHEG-hvq \cite{Frixione:2007nu}, an event
generator for heavy quark production in hadronic collisions at NLO QCD,
where some simple comparisons are presented for top and bottom production.
We also introduce a ``poor man's POWHEG'' for cases where NLO calculations
are available, but only in the traditional phase space slicing approach,
where shower Sudakovs can be used to provide a smoother matching.

One should note that the more sophisticated the description aimed for, 
the higher the price in terms of manpower that goes into the detailed
simulation of a specific process. The point of injecting ME information
is precisely to move away from the universal shower behavior, which 
means that each new process must be considered ``from scratch''.
To provide a sensible first estimate, however, it is useful if 
the shower can be improved to get at least the qualitative behaviour
right for a wide range of processes. In Section 3 we use the 
MadGraph/MadEvent \cite{Alwall:2007st} generator to look at further 
pair-production processes with a jet in the final state. By adding
a Sudakov to the cross-section for real emission of a jet, to approximate
the prescription used by the POWHEG method, we are able to make 
comparisons to the \tsc{Pythia} shower, and find modest changes that
improve agreement significantly.

Specifically, we will address the issue of ``power showers'' vs. 
``wimpy showers''. In \cite{Plehn:2005cq}, the authors compare
two extreme choices for the maximum emission scale of the parton shower;
either the full CM energy of the incoming hadrons (power) or the transverse
mass of the particles produced in the hard collision (wimpy), for both
virtuality- and $\pT$-ordered showers. Their conclusion was that
these options bracketed the matrix-element behaviour for the top and SUSY
production processes studied, but also that the spread of ``predictions''
is large.

One must note that an ultimate goal would be to have a matching scheme
that allows for both a matching to multiple real emissions and to virtual
corrections. Some algorithms have recently been proposed in this
direction \cite{Giele:2007di, Bauer:2008qh, Lavesson:2008ah}, but are not
yet at a stage to be used for serious LHC studies. We will not discuss such
issues further here.

The outline of the article is to study the POWHEG approach and
its relation to \tsc{Pythia} in Section 2, to use MadEvent to gain
an improved understanding of sensible default shower behaviour in
Section 3, and to provide a summary and outlook in Section 4.  

\section{POWHEG}

\subsection{The \tsc{Pythia} merging strategy}

\subsubsection{Evolution equations}
The \tsc{Pythia} parton shower orders final-state radiation (FSR)
emissions in terms of an evolution variable
$Q^2$, such as $m^2$ (previous \tsc{Pythia} versions) or $\pT^2$ (\tsc{Pythia}
nowadays), with an additional energy-sharing variable $z$ in the branching.
For QCD emissions, introducing $t = \ln \left(Q^2/\Lambda_{QCD}^2 \right)$,
the DGLAP evolution equations lead to the probability for a splitting $a
\to bc$
\begin{equation}
\d \mathcal{P} = \sum_{b, c} \frac{\alphas}{2\pi}
P_{a \to bc}(z) \, \d t \, \d z
~,
\end{equation}
where $P_{a \to bc}$ are the DGLAP splitting kernels. This
inclusive quantity can be turned into an exclusive one by requiring that,
for the first (``hardest'') emission, no emissions can have occurred at a
larger $Q^2$. The probability that a branching occurs at $t$ is now given
by
\begin{equation}
\frac{\d \mathcal{P}}{\d t} =
\left( \sum_{b,c} \mathcal{I}_{a \to bc}(t) \right)
\exp \left\{ - \int^{t_{max}}_t \d t' \sum_{b,c}
\mathcal{I}_{a \to bc}(t') \right\}
~,
\label{eq:py_evol}
\end{equation}
with
\[
\mathcal{I}_{a \to bc}(t) =
\int^{z_{max}(t)}_{z_{min}(t)} \d z \,
\frac{\alphas(t)}{2 \pi} \, P_{a \to bc}(z)
~.
\]
The introduction of this Sudakov form factor turns the unnormalised
distribution into a normalised one, i.e.\ with unit integral over the full
phase space. In practice, a lower cutoff, $t_0$, is introduced to keep the 
shower away from soft/collinear regions, which leads to a fraction of 
events with no emissions inside the allowed region.

For initial-state radiation (ISR), the evolution is performed using 
backwards evolution
\cite{Sjostrand:1985xi}, where a given parton $b$ entering a hard
scattering is unresolved into a parton $a$ which preceded it. Here, the
parton distribution functions, reflecting the contents of the incoming
hadron, must be taken into account. Such a change leads to a Sudakov with
the form
\begin{equation}
S_b(x, t_{max}, t) = \exp \left(- \int^{t_{max}}_t \d t' \, \sum_{a,c} \int \d z \,
\frac{\alphas(\pT^2)}{2\pi} \, P_{a \to bc} (z) \,
\frac{x'f_a(x',t')}{xf_b(x,t')} \right)
~.
\label{eq:py_ISR_evol}
\end{equation}

One feature of the above equations is the running renormalisation and 
factorisation 
shower scales, i.e.\ the scales at which $\alphas$ and the PDF's are 
evaluated. For both ISR and FSR, $\alphas$ is evaluated at the $\pT$ scale 
of the emission (the definition of $\pT$ in this context is discussed later 
in Section \ref{sec:py-pt-ordering}). For ISR, the flavour dependent ratio 
of PDF's given in eq.~(\ref{eq:py_ISR_evol}) is evaluated at the selected 
$t$ value which, for the current versions of \tsc{Pythia}, is $\pT^2$.
Thus, the renormalisation and factorisation scales are the same.

\subsubsection{The merging strategy}

Probably the first use of an explicit matching of PS to ME, the 
so-called merging strategy, was introduced to handle the case of 
three-jet events in $\e^+\e^-$ annihilation \cite{Bengtsson:1986hr}. 
An outline is given below, starting from the Born cross section
$\sigma_{\mrm{B}}$ for the lowest-order process $\e^+\e^- \to \gamma^*/Z^0
\to \q\qbar$.

On the ME side, the real-emission cross section 
$\e^+\e^- \to \gamma^*/Z^0 \to \q\qbar\g$ is given, for massless quarks, by
the well-known expression
\begin{equation}
\frac{1}{\sigma_{\mrm{B}}} \, \frac{\d \sigma_{\mrm{R}}}{\d x_1 \d x_2} 
= W^{\mrm{ME}} =  \frac{\alphas}{2\pi} \, \frac{4}{3} \, 
\frac{x_1^2 + x_2^2}{(1 - x_1)(1 - x_2)} ~,
\end{equation}
with $\alphas$ typically evaluated at a scale of $s$, the invariant mass of 
the system.

The DGLAP inclusive $\q \to \q\g$ emission probability in $(Q^2, z)$ space
can be mapped onto the $(x_1, x_2)$ space
\begin{equation}
W^{\mrm{PS}}_{\q} = \frac{\alphas(\pT^2)}{2\pi} \, \frac{4}{3} \,
\frac{1}{Q^2} \frac{1 + z^2}{1 - z} \, 
\frac{\d(Q^2, z)}{\d(x_1, x_2)} ~.
\end{equation} 
The sum of emissions off the $\q$ and $\qbar$ gives
$W^{\mrm{PS}} = W^{\mrm{PS}}_{\q} + W^{\mrm{PS}}_{\qbar}$.
With the addition of a Sudakov form factor, as above, this becomes
\begin{equation}
W^{\mrm{PS}}_{\mrm{corrected}}(Q^2) = W^{\mrm{PS}}(Q^2) \,
\exp \left( - \int_{Q^2}^{Q^2_{\mrm{max}}} \, \d Q'^2 \, 
W^{\mrm{PS}}(Q'^2) \right) ~. 
\label{eq:WPS}
\end{equation} 
For ease of notation we have here omitted the dependence of $W^{\mrm{PS}}$ 
on $z$ and the need of an integration $\d z'$ over a range 
$z'_{\mathrm{min}}(Q'^2) < z' < z'_{\mathrm{max}}(Q'^2)$ in the exponent.

It now so happens that the \tsc{Pythia} shower algorithm covers the full
three-jet phase space and that $W^{\mrm{PS}} > W^{\mrm{ME}}$ everywhere
(so long as the former is true, the latter can always be achieved by a suitable
rescaling). One can therefore use the veto algorithm \cite{Sjostrand:2006za}
to correct down the emission rate. That is, whenever a trial $Q^2$ has 
been selected according to eq.~(\ref{eq:WPS}), the ratio 
$W^{\mrm{ME}} / W^{\mrm{PS}}$ in the chosen phase space point is
the probability that this choice should be retained. If not, the 
evolution is continued downwards from the rejected $Q^2$ scale
(\textit{not} from $Q^2_{\mrm{max}}$). This gives a change from 
eq.~(\ref{eq:WPS}) to 
\begin{equation}
W^{\mrm{PS+ME}}_{\mrm{corrected}}(Q^2) = W^{\mrm{ME}}(Q^2) \,
\exp \left( - \int_{Q^2}^{Q^2_{\mrm{max}}} \, \d Q'^2 \, 
W^{\mrm{ME}}(Q'^2) \right) ~. 
\label{eq:WPSME}
\end{equation} 
Note that, while all explicit dependence on $W^{\mrm{PS}}$ is gone in 
eq.~(\ref{eq:WPSME}), an implicit dependence on the shower remains
in two respects. Firstly, the Sudakov-factor modification of the 
basic ME shape reflects the order in which the shower
algorithm sweeps over phase space, i.e.\ the shower $Q^2$ definition.
Secondly, if the $\alphas$ factors are omitted from the $W^{\mrm{ME}} /
W^{\mrm{PS}}$ reweighting, the $\pT^2$-dependent expression used in the 
shower is retained, instead of the fixed value normal for ME's.

Once the first emission has been considered, an uncorrected shower is 
allowed to continue downwards from the chosen $Q^2$ scale. 
Thus, in this algorithm there is no fixed scale for the transition 
from ME to PS, but a smooth merging of one into the other. This cannot 
give rise to discontinuities in the behaviour at any phase space point, 
except of course at the soft/collinear shower cutoff.

The formalism so far has only considered real emissions. For the 
$\e^+\e^-$ case, however, the cancellation of real (R) and
virtual (V) divergences results in a particularly simple expression
(when integrated over the possible orientations of the event)
\begin{equation}
\sigma_{\mrm{NLO}} = 
\sigma_{\mrm{B}} + \int \d \sigma_{\mrm{R}} + \sigma_{\mrm{V}} 
= \left( 1 + \frac{\alphas}{\pi} \right) \sigma_{\mrm{B}} ~.
\end{equation}
It is therefore trivial to retain eq.~(\ref{eq:WPSME}) as an NLO
prescription for the distribution of events, just by raising the 
cross section associated with each event from $\sigma_{\mrm{B}}$
to $\sigma_{\mrm{NLO}}$. Identifying 
$W^{\mrm{ME}} \, \d Q^2 = \d \sigma_{\mrm{R}} / \sigma_{\mrm{B}}$, we
arrive at the final equation for the differential cross section
\begin{equation}
\d \sigma = \sigma_{\mrm{NLO}} \,
\frac{\d \sigma_{\mrm{R}}}{\sigma_{\mrm{B}}} \, 
\exp \left( - \int_{Q^2}^{Q^2_{\mrm{max}}} \,  
\frac{\d \sigma_{\mrm{R}}(Q'^2)}{\sigma_{\mrm{B}}} \right) ~. 
\label{eq:sigNLO}
\end{equation}

It is thus assumed that the $\mathcal{O}(\alphas)$ 
``new'' part, $\sigma_{\mrm{NLO}} - \sigma_{\mrm{B}}$,
of the total cross section should be associated with the same
radiation function as $\sigma_{\mrm{B}}$. Alternative
choices on this count would only show up in $\mathcal{O}(\alphas^2)$, 
which is beyond the certified accuracy of the algorithm.
Physicswise it is the minimal assumption, relative to having
a different radiation function for the pure NLO part of the
cross section. 

The merging formalism is easy to extend to the case of the emission
of one extra gluon in any $1 \to 2$ decay, and has also been worked 
out and implemented for all such cases within the Standard Model and 
the Minimal Supersymmetric Extension to the Standard Model 
\cite{Norrbin:2000uu}. It can also be applied to ISR for $2 \to 1$ production 
of colour singlet particles, such as $\q\qbar \to \Z^0$ \cite{Miu:1998ju}. 
Here, however, NLO corrections are not so trivial. Experimental practice 
has been to rescale the total cross section to the NLO answer, 
as in eq.~(\ref{eq:sigNLO}), thereby neglecting other potential 
kinematical differences between the LO and NLO answers.  

The merging approach has also been applied to HERWIG
\cite{Seymour:1994df,Corcella:1998rs,Corcella:1999gs}, 
with two main differences. Firstly, since the HERWIG 
shower is ordered in angle rather than hardness, the ME/PS
correction weight must be applied to any emission that is the
hardest so far, rather than only to the first. Secondly, the
HERWIG algorithm leaves holes in the phase space covered by the 
first shower emission, so it becomes necessary also to introduce a 
matching procedure, whereby such holes are filled directly by the 
ME rather than by the PS.

\subsection{The POWHEG strategy}
A general NLO cross section with hadronic incoming states will also contain
remnant (counter) terms from the subtraction of initial-state collinear
singularities, which have already been incorporated into the PDF's. The complete 
differential cross section can therefore be written as the sum of 
contributions from leading order (Born), virtual, real and counter terms
\begin{equation}
\d \sigma = \d \sigma_{\mrm{B}} + \d \sigma_{\mrm{V}} + \,
\d \sigma_{\mrm{R}} - \d \sigma_{\mrm{C}} ~.
\end{equation}

We note that the virtual term has the same $n$-body phase space as the Born
term, while the real and counter terms have an $(n+1)$-body phase space. In
the POWHEG method, the phase space kinematics are factorised in terms of
Born ($v$) and radiation ($r$) kinematic variables, such that the overall
cross section may now be written as
\begin{equation}
\d \sigma = B(v) \, \d \Phi_v + V(v) \, \d \Phi_v + 
R(v,r) \, \d \Phi_v \, \d\Phi_r - C(v,r) \, \d \Phi_v \, \d \Phi_r ~.
\end{equation}
Defining a function that integrates over the radiation variables
\begin{equation}
\bar{B}(v) = B(v) + V(v) + \int d\Phi_r [R(v,r) - C(v,r)] ~,
\label{eqn:POWHEG-NLO}
\end{equation}
once a Born event has been generated, distributed according to 
$\bar{B}(v) \, \d \Phi_v$, the
differential cross section for the hardest emission may be written as
\begin{equation}
\d \sigma = \bar{B}(v)d\Phi_v \left[ \frac{R(v,r)}{B(v)} \exp \left(
-\int_{\pT}
\frac{R(v,r')}{B(v)} d\Phi'_r \right) d\Phi_r \right]
~.
\label{eqn:POWHEG-he}
\end{equation}
The evolution variable of the POWHEG ``shower'' is taken as the kinematical 
$\pT$ of the emission with respect to the parton that branches, identifying 
this highest $\pT$ branching as the hardest. For ISR this is the $\pT$ 
with respect to the beam axis. In this way, the hardest 
radiation is generated according to exact NLO matrix elements, but in a 
probabilistic, exclusive language, which can be directly interfaced to a 
suitable shower program.

This expression shares many features with eq.~(\ref{eq:sigNLO}). The integral
is from some upper scale associated with the Born event, with a lower
cutoff to avoid soft/collinear regions. In both cases it is possible for an 
event to evolve down to this lower cutoff without radiating. Here, however, 
the constant NLO prefactor, $\sigma_{\mrm{NLO}}$, is upgraded to 
$\bar{B}(v)d\Phi_v$, a fully
differential quantity which will encapsulate all kinematical differences
between the LO and NLO answers. It is thus in this term that all the 
sophistication and hard work of a full NLO calculation lies. As in the 
\tsc{Pythia} shower, the radiation is generated with a running $\alphas$ 
expression, evaluated at the $\pT$ scale of the emission.

This formalism has been used for the hadronic production of vector bosons
\cite{Nason:2006hfa,Alioli:2008gx}, heavy quark pairs
\cite{Frixione:2007nw}, single tops \cite{Alioli:2009je} and Higgs bosons
via gluon/vector-boson fusion \cite{Alioli:2008tz,Nason:2009ai}. Codes for
generating all these processes, except for vector boson pairs, is
publicly available. The latest development is the POWHEG BOX; a general
framework for implementing NLO calculations with the POWHEG method
\cite{Alioli:2010xd}.
 
\subsection{The \tsc{Pythia} transverse-momentum-ordered showers}
\label{sec:py-pt-ordering}

The \tsc{Pythia}~8.1 showers are ordered in transverse momentum 
\cite{Sjostrand:2004ef}, both for ISR and for FSR. Also, multiparton 
interactions (MPI) are ordered in $\pT$ \cite{Sjostrand:1987su}.
This allows a picture where MPI, ISR and FSR are interleaved in one 
common sequence of decreasing $\pT$ values \cite{Sjostrand:2007gs}. 
This is most important for MPI and ISR, since they are in direct 
competition for momentum from the beams, while FSR (mainly) 
redistributes momenta between already kicked-out partons.

The interleaving implies that there is one combined evolution equation
\begin{eqnarray}
\frac{\d \mathcal{P}}{\d \pT}  & = &
\left( \frac{\d \mathcal{P}_{\mrm{MPI}}}{\d \pT}  + 
\sum   \frac{\d \mathcal{P}_{\mrm{ISR}}}{\d \pT}  +
\sum   \frac{\d \mathcal{P}_{\mrm{FSR}}}{\d \pT} \right)
\nonumber \\ 
& \times & \exp \left( - \int_{\pT}^{p_{\perp\mrm{max}}} 
\left( \frac{\d \mathcal{P}_{\mrm{MPI}}}{\d \pT'}  + 
\sum   \frac{\d \mathcal{P}_{\mrm{ISR}}}{\d \pT'}  +
\sum   \frac{\d \mathcal{P}_{\mrm{FSR}}}{\d \pT'} \right) 
\d \pT' \right)
\label{eq:combinedevol}
\end{eqnarray}
that probabilistically determines what the next step will be.
Here the ISR sum runs over all incoming partons, two per
already produced MPI, the FSR sum runs over all outgoing partons, 
and $p_{\perp\mrm{max}}$ is the $\pT$ of the previous step.
Starting from a single hard interaction, eq.~(\ref{eq:combinedevol})
can be used repeatedly to construct a complete parton-level event 
of arbitrary complexity. Recently also rescattering has been 
included as a further (optional) component of the MPI framework 
\cite{Corke:2009tk}.

The decreasing $\pT$ scale can be viewed as an evolution towards
increasing resolution; given that the event has a particular structure
when activity above some $\pT$ scale is resolved, how might that 
picture change when the resolution cutoff is reduced by some infinitesimal
$\d \pT$? It does not have a simple interpretation in absolute time;
all the MPI occur essentially simultaneously (in a simpleminded
picture where the protons have been Lorentz contracted to pancakes),
while ISR stretches backwards in time (and is handled by backwards 
evolution \cite{Sjostrand:1985xi}) and FSR forwards in time. 
The closest would be to view eq.~(\ref{eq:combinedevol}) as an 
evolution towards increasing formation time.

For the following, a relevant aspect is that the $\pT$ definition is 
not exactly the same for MPI, ISR and FSR in \tsc{Pythia}. For an MPI the $\pT$ 
is the expected one; the transverse momentum of the two scattered partons 
in a $2 \to 2$ process, defined in a Lorentz frame where the two 
incoming beams are back-to-back. To understand why problems arise
for ISR or FSR, consider a $\q \to \q\g$ branching, 
where the $\pT$ of the emitted gluon is defined with respect to the 
direction of the initial quark. The $\pT$ as a function of the 
gluon emission angle $\theta$ increases up till $90^{\circ}$, but then 
decreases again, $\pT \to 0$ for $\theta \to 180^{\circ}$. Thus, an 
ordering in such a $\pT$ would classify a $\sim 180^{\circ}$ emission 
as collinear and occurring late in the evolution, although it would 
involve a more off-shell propagator than an emission at $90^{\circ}$. 
It could also erroneously associate a $1/\pT^2$ divergence with the 
$\theta \to 180^{\circ}$ limit. Therefore it is natural to choose an 
evolution variable that does not turn over at $90^{\circ}$. 

To this end, consider a branching $a \to b c$ (e.g.\ $\q \to \q \g$), 
where $z$ is defined as the lightcone (LC) momentum along the $a$ axis 
that $b$ takes. Then
\begin{equation}
p_{\perp\mrm{LC}}^2 = z(1-z)m_a^2 - (1-z)m_b^2 - z m_c^2 ~,
\end{equation}
and this equation can be used as inspiration to define evolution variables
\begin{eqnarray}
\mrm{ISR:} & \pTevol^2 = (1-z) Q^2 & \mrm{with}~
m_b^2 = -Q^2 ~\mrm{and}~ m_a^2 = m_c^2 = 0 ~, \\
\mrm{FSR:} & \pTevol^2 = z (1-z) Q^2 & \mrm{with}~
m_a^2 = \phantom{-}Q^2 ~\mrm{and}~ m_b^2 = m_c^2 = 0 ~, 
\end{eqnarray}
which are monotonous functions of the virtuality $Q^2$. However, once a 
branching has been found and the kinematics is to be reconstructed, 
the $Q^2$ interpretation is retained, but $z$ is now replaced by a Lorentz 
invariant definition. For ISR this is chosen to be 
$z = m^2_{br}/m^2_{ar}$, where $r$ is the ``recoiler'', i.e.\ the 
incoming parton from the other side of the subcollision. The actual 
$\pT$ of $b$ and $c$ then becomes 
\begin{equation}
p_{\perp b,c}^2 = (1-z)Q^2 - \frac{Q^4}{m_{ar}^2} = 
\pTevol^2 - \frac{\pTevol^4}{p_{\perp\mrm{evol,max}}^2} ~,
\label{eq:pTbc}
\end{equation}
where $p_{\perp\mrm{evol,max}}^2 = (1-z) Q_{\mrm{max}}^2 = 
(1-z)^2 m_{ar}^2$. Instead, for FSR, $z = E_b/E_a$ in the rest frame of
$a$ and its colour-connected recoiler. The $p_{\perp_{b,c}}^2$ expression
in this case becomes somewhat lengthier than eq.~(\ref{eq:pTbc}), but 
shares the same physical properties; at small values $\pTevol^2$
and $p_{\perp b,c}^2$ follow suit, and so correspond to identical 
$1/\pT^2$ singular behaviours, but the latter then turns around and 
vanishes when $\pTevol^2$ approaches the kinematical limit. 

\subsection{Interfacing POWHEG-hvq to \tsc{Pythia}}

In this study, the interface from POWHEG-hvq to \tsc{Pythia} was performed 
using the Les Houches Event File (LHEF) \cite{Alwall:2006yp} format. Both 
programs come with a number of PDF sets available for selection and 
\tsc{Pythia} gives access to external PDF sets through the LHAPDF library 
\cite{Whalley:2005nh}. For consistency, the CTEQ6L \cite{Pumplin:2002vw} 
PDF set has been used throughout this study, as this set is available in 
all three programs used; no other differences in the PDF set itself have 
been taken into account. One caveat is the associated $\alphas$ running 
expression for a given PDF set. \tsc{Pythia} by default will use the first 
order expression, and allow $\alphas(M_Z^2)$ to be set independently for 
spacelike and timelike showers. Here, $\alphas(M_Z^2)$ was set to match the 
one used in the PDF set, but with first order running.

Historically the LHA conventions \cite{Boos:2001cv}, used by LHEF, only 
encompass one event ``scale'', specified to be the factorisation one.
The $\alphas$ and $\alphaem$ values at the renormalisation scale are 
also supplied, but not that scale itself, and there is no provision for
shower-matching scales. A proposal for an extended standard that 
would store more scales has recently been presented \cite{Butterworth:2010ym}.
When reading an LHEF into a shower program, the question then is: at what
scale should the subsequent shower evolution begin?

In \tsc{Pythia}, such choices need to be made for ISR and FSR separately.
The basic principle should be to avoid doublecounting to the largest
extent possible. With LHEF input it would be natural to let the 
showers begin at the factorisation scale, the only known one, and 
then proceed downwards. This is an allowed choice both for ISR 
and FSR, but for ISR another possibility is default. Here, events 
are split into two kinds, based on the absence or presence of particles 
that the shower can produce, i.e.\ $\d$, $\u$, $\s$, $\c$, $\b$, $\g$ 
and $\gamma$. If the LHEF final state contains no such particles then 
the shower can populate the full phase space without any risk of 
doublecounting, which should give more realistic event shapes. This 
would be the case for many $\W/\Z$, top, Higgs and New Physics processes. 
If, on the other hand, the final state does contain particles that could be
produced in the shower, then doublecounting would be more likely 
than not, and the factorisation scale again becomes the only reasonable 
choice. 

This more flexible attitude works well if the ME program does not mix
different topologies, but breaks down if, say, both $\t \tbar$ and
$\t \tbar \g / \t \tbar \q$ events are supplied, with the latter intended
to correspond to the fraction of $\t \tbar$ events with 
an extra emission above the factorisation scale. When such a mixing is
present, the showers off $\t \tbar$ should not be allowed to populate the
whole phase space. Thus, we conclude that interfacing to an LHEF cannot be
done completely automatically, but must be made with some knowledge of
which rules were used to produce the LHEF.

Therefore, to provide a consistent interface to POWHEG, we must both consider 
its generation strategy and the information it stores in the LHEF.
In events with an emission above the lower cutoff scale $\pTmin$, 
POWHEG chooses the factorisation scale to be the $\pT$ of the emission,
$\pTPOW = \pT$ (defined with respect to the beam axis in the case of
POWHEG-hvq).

When the POWHEG ``shower'' reaches $\pTmin$ without any emissions, the
factorisation scale is instead set equal to this $\pTmin$, $\pTPOW =
\pTmin$. Since $\pTmin$ normally is rather small, of the order of $1\GeV$,
the fraction of no-emission events is also small.

The LHEF choice of storing $\alphas$ and $\alphaem$ rather than the 
renormalisation scale makes sense for traditional ME calculations, 
where typically one fixed scale is used. The POWHEG case is somewhat 
special, as it can have different renormalisation scales for the Born 
level process and for the subsequent radiative emission.

For the case where POWHEG has already generated an emission, the obvious
choice is to begin the evolution at the factorisation scale $\pTPOW$, 
such that the shower will not generate harder emissions. However, there 
are then two potential complications: first, the mismatch between 
$\pTPOW$ and the lightcone-inspired $\pTevol$ scales of \tsc{Pythia}, and
second, specifically for FSR, that $\pT$ is defined with respect to 
the direction of the emitting parton rather than to the beam axis. 
For ISR, eq.~(\ref{eq:pTbc}) shows that $\pT < \pTevol$, such that starting 
the shower from $\pTevol = \pTPOW$ will lead to a small area of phase 
space not being covered. An FSR emission, on the other hand, may have 
a small $\pTevol$ with respect to the emitting parton but still a
$\pT > \pTPOW$ with respect to the beam axis.

A simple solution to both these problems is instead to begin the shower
at the largest possible scale, and then veto any emissions with a 
kinematic $\pT > \pTPOW$. If we consider the first shower emission, 
the multiplicative nature of the no-emission probability ensures that 
the emission rates below $\pTPOW$ will be correct, i.e.\ unaffected by the
vetoes above it. The picture is slightly less clear for subsequent
emissions; having accepted one shower emission below $\pTPOW$, it is still
possible for a later emission to be above it, since the first emission may
well have had $\pTevol > \pTPOW$. The probability of such an occurrence is
small, and effects formally of NNLO character, unenhanced by any large
logarithms. They mainly show up for low-$\pT$ first emissions, where their
importance on the event as a whole is less, but still nonzero.  Another
NNLO issue is that recoil effects from one emission can shift the $\pT$ of
the previous ones, along with the hard process itself, either to lower or
higher values.

The current POWHEG-hvq generator uses a second order running $\alphas$
expression, but with a $\Lambda$ fixed at $n_f = 5$. Although slightly
inconsistent, this only leads to changes beneath the Bottom and Charm
scales. The $\Lambda$ value is taken from a selected PDF set and is
modified as in \cite{Catani:1990rr}. In the LHEF output file, all incoming
and radiated partons are massless and the values of the couplings,
$\alphas$ and $\alphaem$, are set to zero in all events.

\begin{figure}
\centerline{
\includegraphics[angle=270,scale=0.63]{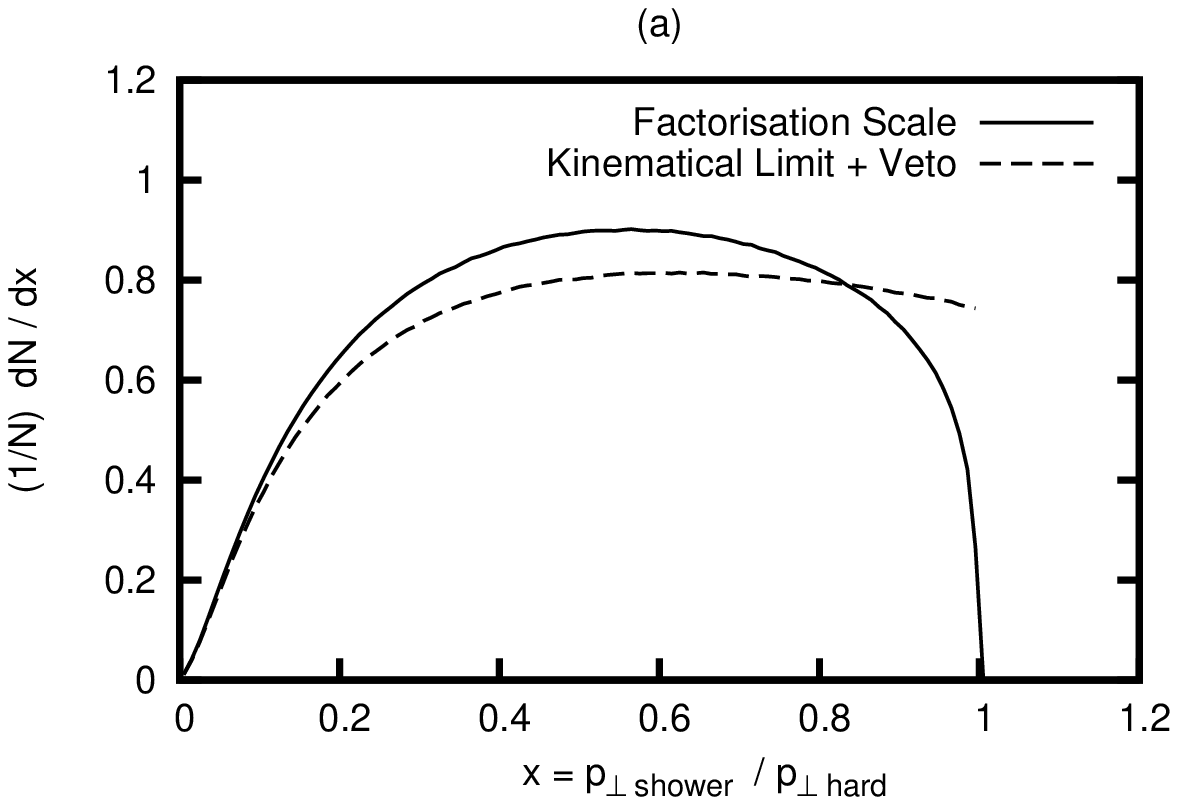}
\includegraphics[angle=270,scale=0.63]{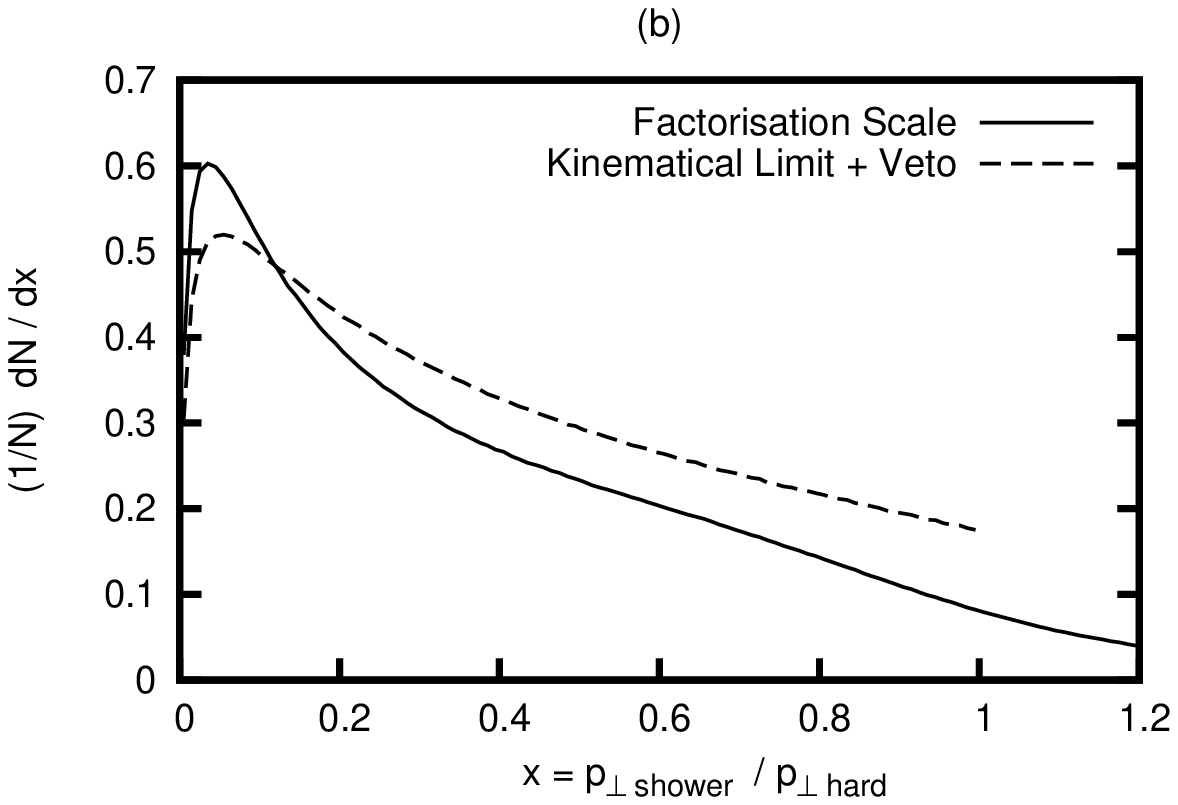}
}
\caption{Ratio of the kinematic $\pT$ of the first shower emission to the
POWHEG emission, where the shower emission is (a) ISR or (b) FSR. In both
cases, the results are shown when starting the shower at the factorisation
scale and when starting the shower at the kinematical limit and vetoing
above the POWHEG scale
\label{fig:ph-py-ttbar}
}
\end{figure}

To quantify how well the proposed interfacing works, we begin with top pair
production ($m_{\t} = 171\GeV$), where all results are generated at LHC
energies ($\p\p$, $\sqrt{s} = 14\TeV$). In this case, the
number of light flavours, which defines the content of the proton and the
allowed radiation flavours, goes up to and includes the bottom quark
($n_l = 5$). To study the effect of the different shower starting scales,
we examine the ratio of the first $\pT$ in the shower to the $\pT$ of
the POWHEG emission (where the shower $\pT$ value is taken directly after
the emission). This is shown in Fig.~\ref{fig:ph-py-ttbar}, split
into contributions from (a) ISR and (b) FSR. For ISR, we note that the
ratios do not become larger than unity, but that when starting the shower
at the factorisation scale, there is a region close to 
$p_{\perp\mrm{shower}}/p_{\perp\mrm{hard}} = 1$ where the phase space is not 
completely filled. This gap is filled when starting the shower at the 
kinematical limit and vetoing emissions above $\pTPOW =
p_{\perp\mrm{hard}}$; that is, the shower is running at ``full steam'' when
it reaches the emission $\pT$ threshold. That the curve happens to be so
flat near the endpoint is a coincidence, related to a cancellation between
the blowup of the na\"{\i}ve emission rate for smaller $\pT$, with a Sudakov
damping in the same limit, helped along by many events having a small
$p_{\perp\mrm{hard}}$ in the first place. It does not happen for FSR, where
the large top mass reduces radiation to a lower level overall. For FSR,
when starting at the factorisation scale, there is a tail
beyond unity, as discussed previously, while this no longer happens with
the veto scheme in place. Note that the FSR rate below $\pTPOW$ does come
up, however, since emitted partons now are allowed at a larger separation
from their mother parton so long as they are still at small $\pT$ with
respect to the beam. As already noted, the corrections are of higher order,
but their inclusion is worthwhile for overall consistency.

\begin{figure}
\centering 
\includegraphics[angle=270,scale=0.8]{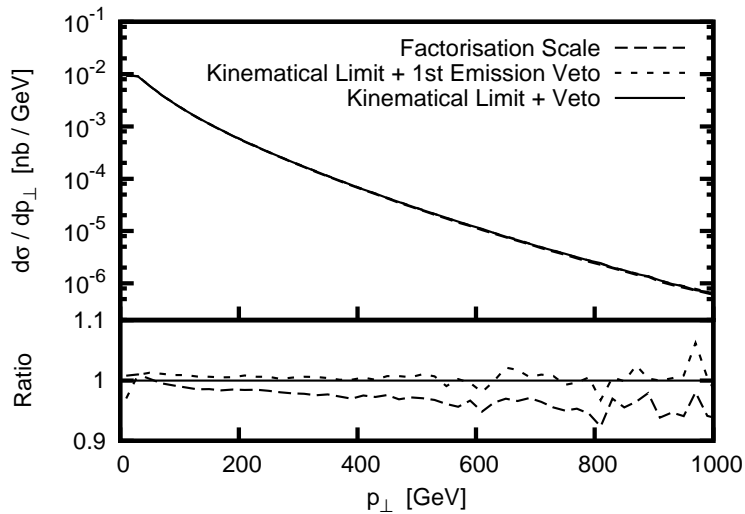}
\caption{Final $\pT$ of the top pair for three different cases (see text).
In the upper plot, it is difficult to distinguish the different curves, but
the results are clearer to see in the lower plot, which shows the ratio to
the ``Kinematical Limit + Veto'' result
\label{fig:ph-py-ttbar-pt}
}
\end{figure}
 
\begin{figure}
\centerline{
\includegraphics[angle=270,scale=0.63]{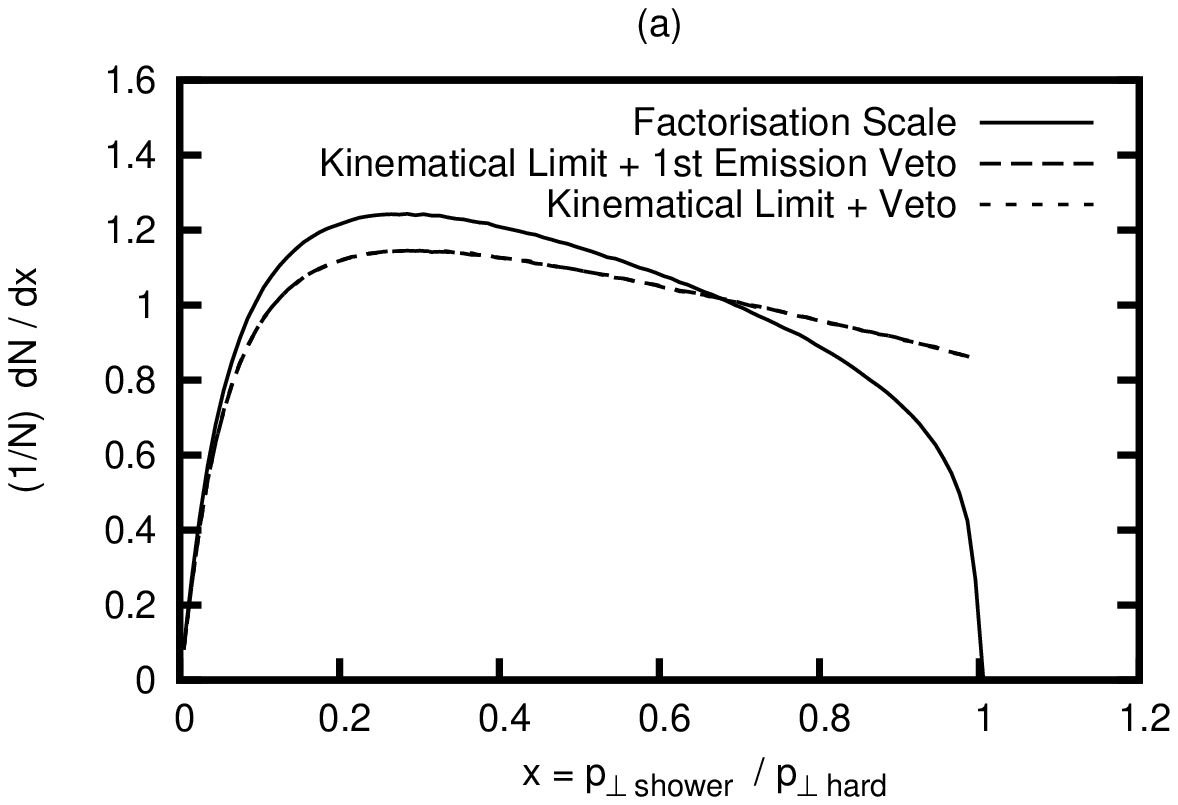}
\includegraphics[angle=270,scale=0.63]{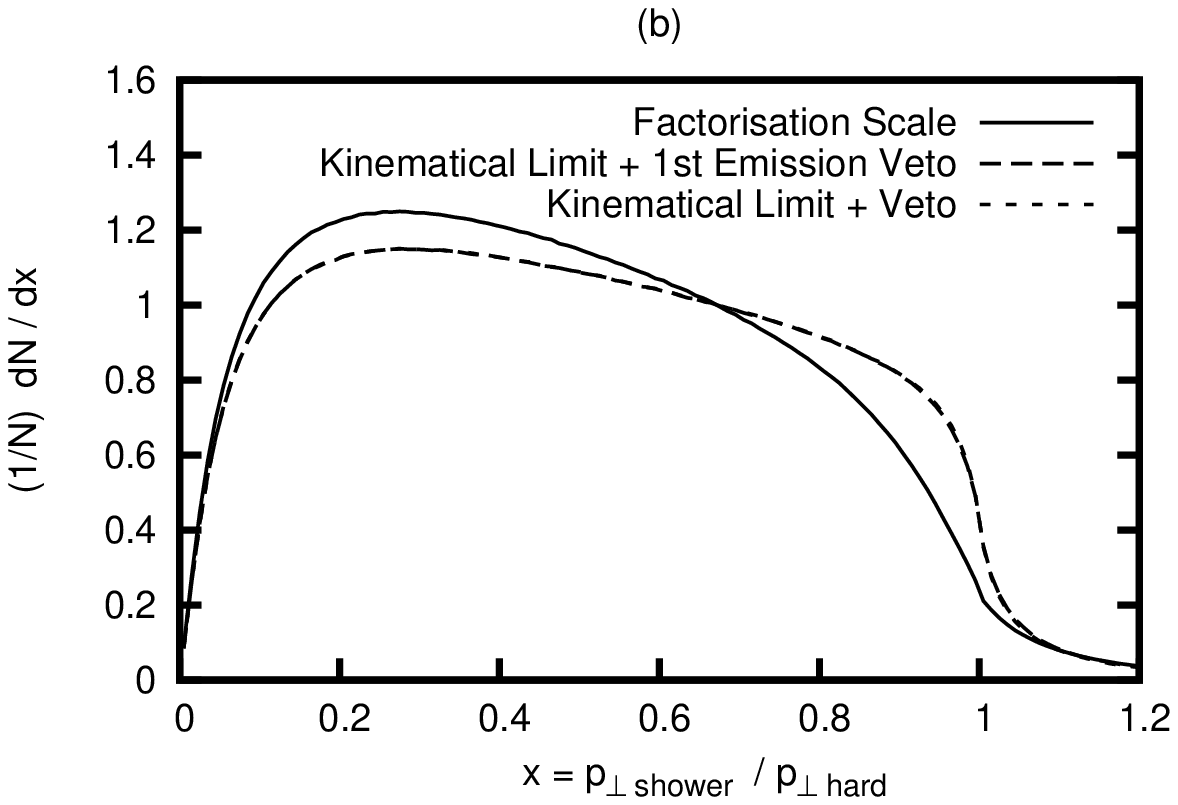}
}
\caption{Ratio of the kinematic $\pT$ of the first shower emission to the
POWHEG emission for ISR only. (a) Shows the results immediately after the
first emission, while (b) shows the results after the full shower
evolution. In both cases, the ``Kinematical Limit + 1st Emission Veto'' and
``Kinematical Limit + Veto'' curves lie on top of each other
\label{fig:ph-py-ttbar-ISRonly}
}
\end{figure}

We move on to study the effect of shower emissions beyond the first. As
discussed previously, after a first allowed shower
emission below $\pTPOW$, a subsequent emission may be generated above $\pTPOW$. 
To examine this effect, three different cases are considered: (1) the showers are
started at the factorisation scale, (2) the showers are started at the
kinematical limit and only the first shower emission is vetoed, and (3) the
showers are started at the kinematical limit and all emissions above
$\pTPOW$ are vetoed. As a measure of this effect, one may use the final $\pT$
of the top pair, which is shown in Fig.~\ref{fig:ph-py-ttbar-pt}
for the three different cases. At first glance, all
three approaches appear to give similar results, but the ratio
plot reveals that the difference between the factorisation scale and the
veto schemes is around 10\%, while there is little difference between the
two different veto schemes. Instead, in Fig.~\ref{fig:ph-py-ttbar-ISRonly},
we study the smearing of the first shower emission $\pT$ due to
subsequent emissions. In this case only ISR is generated, so there is no
ambiguity in picking out the first shower emission in the final event,
and again, we show the ratio of the first shower emission to the
POWHEG one, $p_{\perp\mrm{shower}}/p_{\perp\mrm{hard}}$, (a) immediately
after the first emission and (b) after the shower evolution has finished.
Fig.~\ref{fig:ph-py-ttbar-ISRonly}a shows the same features as
Fig.~\ref{fig:ph-py-ttbar}, although with different normalisation due to
the lack of FSR. In Fig.~\ref{fig:ph-py-ttbar-ISRonly}b, the
first shower emission $\pT$ is smeared by small amounts due to subsequent
emissions, but again the difference between vetoing just the
first emission and all emissions is negligible.

\begin{figure}
\centerline{
\includegraphics[angle=270,scale=0.63]{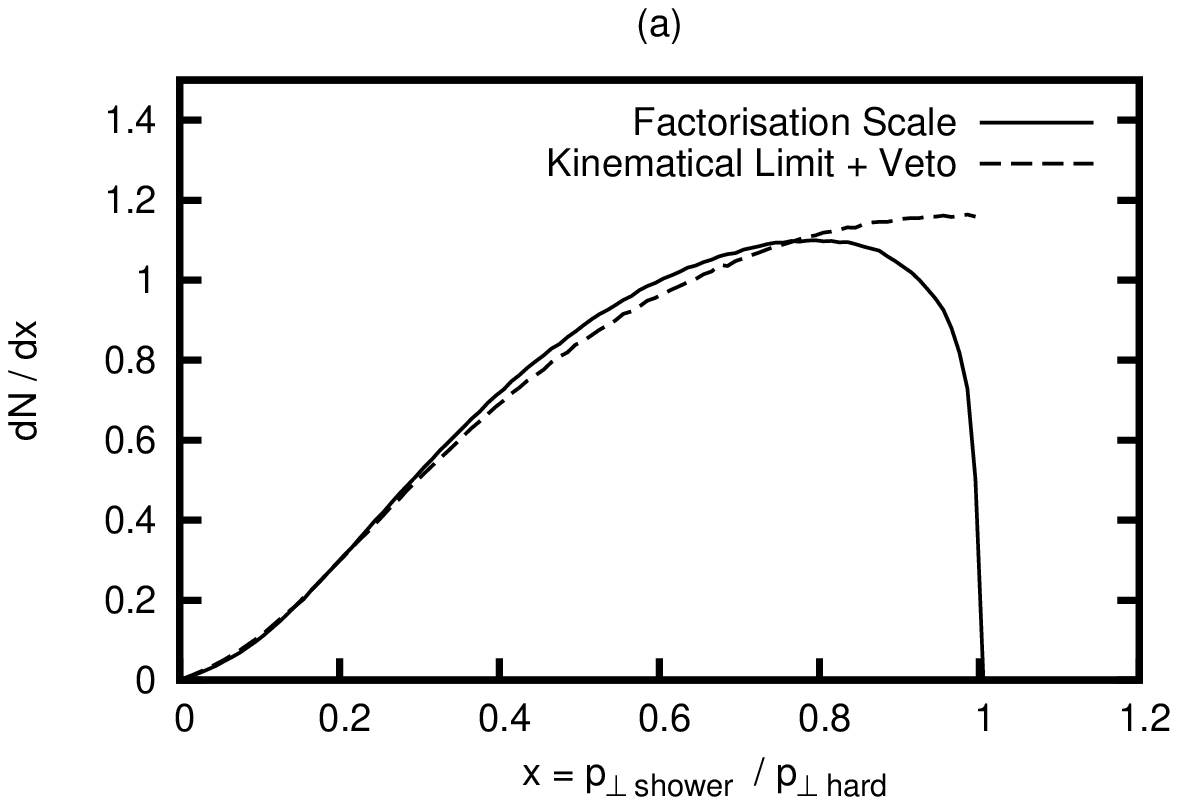}
\includegraphics[angle=270,scale=0.63]{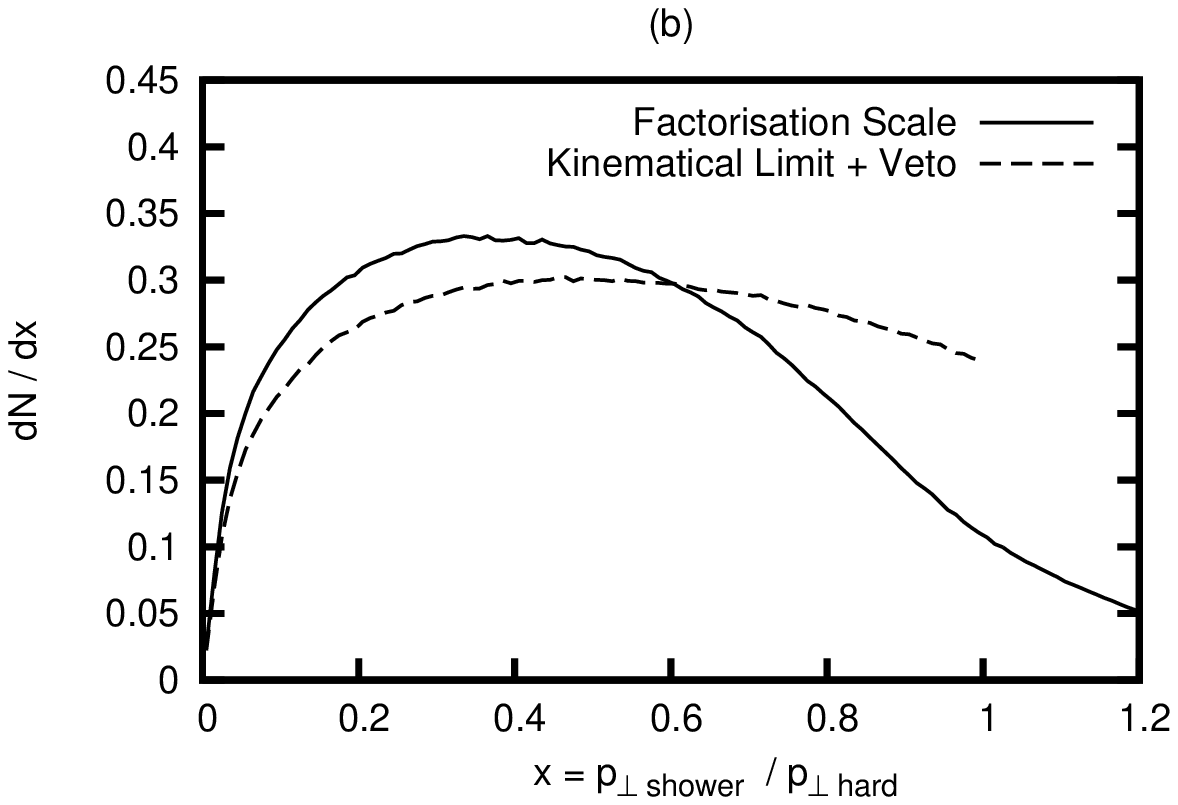}
}
\caption{Ratio of the kinematic $\pT$ of the first shower emission to the
POWHEG emission for bottom pair production, where the shower emission is
(a) ISR or (b) FSR
\label{fig:ph-py-bbar}
}
\end{figure}

For bottom production ($m_{\b} = 4.8\GeV$), the number of light flavours is
now set to four ($n_l = 4$), meaning that there will be no incoming or
radiated $\b$ quarks. This is different from the default
\textsc{Pythia} settings, where $\b$ quarks are both taken from the beam
and allowed to be created in radiative emissions. The results are shown in
Fig.~\ref{fig:ph-py-bbar}, again split into contributions from (a) ISR and
(b) FSR. Although the overall pattern of radiation is different, due to the
smaller bottom mass, the details with respect to the interface with
\tsc{Pythia} show the same features as Fig.~\ref{fig:ph-py-ttbar}.

\subsection{Poor Man's POWHEG}

The POWHEG approach is a very powerful tool for NLO studies.
Unfortunately, POWHEG-based implementations are not available
for all relevant processes. Many older NLO calculations are
instead available in terms of phase space slicing results
(see e.g.\ \cite{Giele:1993dj}).
That is, the $(n+1)$-body cross section is based on the pure
real-emission cross section $\d\sigma_{\mrm{R}}$, without any Sudakov
corrections. Unresolved emissions, virtual corrections and
counterterms are lumped together with the Born cross section
to provide the $n$-body cross section. The absence of a Sudakov
factor means that soft/collinear divergences in $\d\sigma_{\mrm{R}}$
are not dampened, thus requiring a reasonably large cutoff scale
to stay away from the region where $\sigma_n$ would turn negative.
On the other hand, $\d\sigma_{\mrm{R}}$ should be used where it can.
Therefore $\sigma_n$ and $\sigma_{n+1}$ typically are chosen to be
of the same order, while $\sigma_n \ll \sigma_{n+1}$ in POWHEG,
where a much smaller cutoff can be used without any risk of
inconsistencies.

In order to make these older NLO codes also useful to the
experimental community they need to be interfaced to event generators,
for further showering, MPI and hadronisation. It is then convenient
to add Sudakovs to $\d\sigma_{\mrm{R}}$, to bring it closer to the POWHEG
approach. This would also smoothen out the transition between $(n+1)$-
and $n$-body phase space, in the spirit of modern-day leading-order
matching procedures, already mentioned in the introduction.

As a practical example we mention the \textsc{Wgamma\_nlo} package for
$\W\gamma$ production to NLO \cite{Baur:1993ir}, with up to one
additional quark or gluon in the final state, and allowing for
anomalous $\W\W\gamma$ couplings. Here an implementation of the
principles to be presented is already available \cite{Majumder:2010ax}.

Assume a lowest-order process that does not contain any coloured
particles in the final state. The NLO processes thus include one
quark or gluon extra, with divergences when its $\pT \to 0$. A scale
$\pTmin$ is used to separate the $(n+1)$- and $n$-body phase space.
An $(n+1)$-body state is characterised by the transverse momentum of the
quark or gluon, $\pT^{\mrm{ME}}$, with $\pT^{\mrm{ME}} > \pTmin$. We now
want to include the Sudakov to express that $\pT^{\mrm{ME}}$ is the hardest
jet of the event, i.e.\ that there are no jets at a larger scale.

In the CKKW-L \cite{Lonnblad:2001iq} approach a fictitious trial shower
is used to provide this. The better this shower attaches to the correct
ME behaviour, the more accurate the Sudakov will be (thus, an advantage
to having a more accurate default behaviour of the shower, as we strive
for in this article). The point should not be overstressed, however;
at large $\pT$ the Sudakov suppression is negligible anyway, and at
small $\pT$ the universal behaviour should dominate.

The $(n + 1)$ topology has to be projected down to the core $n$ process,
without the extra parton, to provide the starting point for the trial
shower. This requires a choice of which side the emission occurs on.
Sometimes flavours allow only one possibility;
for $\u \g \to \W^+ \gamma \d$ the hard subprocess must be
$\u \dbar \to \W^+ \gamma$ since the gluon cannot couple to the
$\W^+ \gamma$ state. For $\u \dbar \to \W^+ \gamma \g$ the emission
could be on either side, and the relative probability for a shower
emission is related to the respective $1/p_{\perp\mrm{evol}}^2$. Since
$z = m^2_n / m^2_{n+1}$ is independent of emission side, it follows
that $p_{\perp\mrm{evol}}^2 \propto Q^2 \propto 1 \mp \cos\hat{\theta}$,
where $\hat{\theta}$ is the angle of the gluon in the rest frame of the
subcollision. The constancy of $z$ also implies that the splitting
kernel values are the same on both sides, and so give no net contribution.
In another case, where the core process would be different depending
on the side of emission, the weight of the respective splitting kernel
would have to be taken into account, but still at the same $z$.
In summary, the relative probability for a gluon emission from side 1
is $Q_1^{-2} / (Q_1^{-2} + Q_2^{-2}) = Q_2^2 / (Q_1^2 + Q_2^2) =
(1 + \cos\hat{\theta})/2$. The choice of side determines a new
$x' = x z$ for it, while $x$ on the other side is unchanged.
A combined rotation and boost can bring the $n$ particles to this
new frame.

From there, the fictitious shower is allowed to generate the first/hardest
emission at a scale $\pT^{\mrm{PS}}$. The probability for
$\pT^{\mrm{ME}} > \pT^{\mrm{PS}}$ is precisely the desired Sudakov,
in the CKKW-L spirit. That is
\begin{enumerate}[(i)]
\setlength{\itemsep}{-10pt}
\item if $\pT^{\mrm{ME}} > \pT^{\mrm{PS}}$ the original
$n + 1$-body topology is retained;\\
\item else the projected $n$-body topology is selected.
\end{enumerate}
In case (i), a normal shower can then be started up from
$\pT^{\mrm{ME}}$ and downwards; that is, further jets may be generated
above the $\pTmin$ scale, to give $(n + 2)$-body topologies etc.
The events in case (ii) are to be lumped together with
the ones already originally classified as $n$-body, and allowed
to shower from the $\pTmin$ scale downwards.

In this approach, there is not guaranteed to be a smooth matching at
$\pTmin$, at least for the $\pT$ scale of the hardest emission.
The hope is that this step will be smeared out by subsequent showers
and hadronization. Also note that, unlike POWHEG defined by
eq.(\ref{eqn:POWHEG-he}), but in line with MC@NLO, the high-$\pT$ tail is
defined by the LO $(n + 1)$ expression, without any $\bar{B}(v) / B(v)$ ``K
factor''.

\section{POWHEG without NLO}

\subsection{Correction to the \tsc{Pythia} shower}
\label{sec:py-correction}

As will be shown in Section \ref{sec:top} for top pair production, an
evolution with $Q_{\mrm{max}}^2 = s$ (power) overestimates the
high-$\pT$ tail while $Q_{\mrm{max}}^2 = m_{\perp t}^2$ (wimpy)
underestimates it, cf. \cite{Plehn:2005cq}. This is not so surprising, as
follows. Let us recall that the matrix element for a QCD process such as
$\g \g \to \g \g \g$ scattering roughly behaves like
\cite{Berends:1981rb,Andersson:1995jt}
\begin{equation}
\d\sigma \sim \frac{\d p_{\perp 1}^2}{p_{\perp 1}^2} 
\frac{\d p_{\perp 2}^2}{p_{\perp 2}^2} \frac{\d p_{\perp 3}^2}{p_{\perp 3}^2}
\, \delta^{(2)}\left( \mbf{p}_{\perp 1} + \mbf{p}_{\perp 2} + \mbf{p}_{\perp 3}
\right) ~, 
\end{equation}
where the $p_{\perp i}$ are the transverse momenta of the three outgoing 
gluons. In the limit $p_{\perp 3} \ll p_{\perp 1}$, where 
$|\mbf{p}_{\perp 1}| \approx |\mbf{p}_{\perp 2}|$, this reduces to 
$1/p_{\perp 1}^4 \, p_{\perp 3}^2$, i.e.\ the hard interaction behaves 
like $\d p_{\perp 1}^2 / p_{\perp 1}^4$ and the subsequent shower
emission of an additional gluon like $\d p_{\perp 3}^2 / p_{\perp 3}^2$.
Put another way, for a fixed $p_{\perp 1}$, you may distinguish a
low-$p_{\perp 3}$ region with a fall-off like 
$\d p_{\perp 3}^2 / p_{\perp 3}^2$ and a high-$p_{\perp 3}$ region where
the fall-off instead is like $\d p_{\perp 3}^2 / p_{\perp 3}^4$,
with a smooth transition when $p_{\perp 3} \approx p_{\perp 1}$.  
In practice you would not want to simulate the process like this, 
of course, but reserve the hardest propagator to be described by 
matrix elements, which also would avoid doublecounting problems.

Obviously the picture is not equivalent for the $\g \g \to \t \tbar \g$,
but let us apply a similar reasoning to the extra gluon in this case.
For small $p_{\perp \g}$ the
picture of an ISR branching $\g \to \g \g$ followed by a hard
process $\g \g \to \t \tbar$ ought to be a valid approximation, and 
so we expect a $\d p_{\perp \g}^2 / p_{\perp \g}^2$ falloff. At large
$p_{\perp \g}$, on the other hand, it would make more sense to 
think in terms of an ISR branching $\g \to \t \tbar$ followed by a
hard process $\g \t \to \g \t$, and thus a shape more like 
$\d p_{\perp \g}^2 / p_{\perp \g}^4$. Now, we don't simulate the latter 
kind of hard processes, e.g.\ because the top is so heavy that a top
PDF is not a fruitful approximation, and so we would like to obtain 
these high-$p_{\perp \g}$ configurations in the context of the simulation 
of the $\g \g \to \t \tbar$ process. This leads to an ansatz of the form
\begin{equation}
\frac{\d\mathcal{P}_{\mrm{ISR}}}{\d \pT^2} \propto
\frac{1}{\pT^2} \, \frac{k^2 M^2}{k^2 M^2 + \pT^2} ~,
\label{eq:pTdamp}
\end{equation}
where $M^2$ is a reasonable scale to associate with the hard $2 \to 2$
process (discussed further below) and $k^2$ is a fudge factor 
of order unity, parameterising at what scale the transition from a
$1/\pT^2$ to a $1/\pT^4$ behaviour occurs.  

How generic would such an ansatz be? First of all, for QCD processes
involving light quarks and gluons, the showers should be cut off at
(around) the scale of the hard process, or else one would doublecount,
since in this case all possible $2 \to 2$ hard subprocesses one could
construct out of a $2 \to 3$ process are already being simulated. Secondly,
for $2 \to 1$ production of a colour singlet particle, the $\d \pT^2 /
\pT^2$ ansatz works well up to the kinematical limit \cite{Miu:1998ju}.
A reasonable assumption is that this generalises to processes where two or
more colour singlet particles are produced in the core process, while a
shape similar to eq.~(\ref{eq:pTdamp}) should occur in processes that
involve one or several coloured particles in the final state.

The argument for such a difference is one of colour coherence; with colour
charge in both the initial and the final state one expects a destructive
interference between ISR and FSR emissions that limits the radiation
\cite{Ellis:1986bv}, while no such interference occurs with colours only in
the initial state. One can then argue exactly what the scale $M^2$
appearing in eq.~(\ref{eq:pTdamp}) should be; for pair production of
coloured particles, the factorisation or renormalisation scale should be a
reasonable choice, but when considering the production of a mixed
coloured/non-coloured final state, by the coherence argument, it is
primarily the coloured particles that should play a role in this scale. We
note that the default choice for internal \tsc{Pythia} $2 \to 2$ processes
is to set the renormalisation scale equal to the geometric mean of the
squared transverse masses of the two outgoing particles, while the
factorisation scale is set to the smaller of the squared transverse masses
of the two outgoing particles. In the next sections we will check how well
the ansatz of eq.~(\ref{eq:pTdamp}) fares for a number of different
processes.

\subsection{MadEvent}

\subsubsection{Hard emission probability}
\label{sec:me-method}

In order to study processes for which an implementation of
POWHEG is not yet available, we turn to the MadEvent matrix
element generator. In eq.~(\ref{eqn:POWHEG-he}) we see that the POWHEG
Sudakov, used to generate hard emissions, contains only Born and real
terms; all NLO corrections are contained in a separate prefactor.

We therefore use MadEvent to generate events with an extra jet in the final
state in order to extract the cross section for real jet emission, $\d
\sigma_R$.  An approximate ``POWHEG'' style probability for emission is
then formed by normalising to the overall lowest order (Born) cross section
(generated by simulating the corresponding $2 \to 2$ process in MadEvent)
and adding a Sudakov. This leads to a probability distribution described
by
\begin{equation}
\d \mathcal{P} =
\frac{\d \sigma_{\mrm{R}}}{\sigma_{\mrm{B}}}
\exp \left( - \int_{\pT}^{{\pT}_{max}}
\frac{\d \sigma_{\mrm{R}}}{\sigma_{\mrm{B}}} \right)
~,
\end{equation}
where the $\pT$ integral begins at the kinematical limit. As noted 
previously, the NLO prefactor can lead to kinematical differences in the 
resulting distribution. Under the assumption that such kinematical 
differences are small (to be addressed further in the case of top 
pair production), a qualitative comparison to the \tsc{Pythia} shower can
be made. The effect of the Sudakov will be most visible in the low-$\pT$
region, where the ME $\d \sigma_R$ calculation is divergent. Here the
integrand will blow up and the Sudakov will make the distribution turn
over, such that it will have a unit integral (up to cutoff effects). Given
that the low-$\pT$ region is where the parton shower should be most
accurate, we expect the turnover in the Sudakov-modified MadEvent
distribution to roughly correspond with the \tsc{Pythia} distribution.

All Standard Model masses in MadEvent are set equal to the default
\textsc{Pythia} values. Variations in particle widths between MadEvent and
\textsc{Pythia} are not expected to play a large role and are therefore
neglected. The definition of both the proton and jets includes bottom
quarks, as would be included in the \textsc{Pythia} default settings. In
what follows, we will only consider heavy final states (top mass and
higher), such that we can neglect FSR and only consider ISR. In this case,
there are only soft and collinear singularities from the branching of
incoming partons. A low-$\pT$ cutoff of $2\GeV$ is introduced on
real-jet emission, such that these divergences are avoided. No further cuts
are applied.

\subsubsection{Scale corrections}

In generating events with MadEvent, there are again some different choices
available relating to running $\alphas$ expressions and PDF's. The choice
of PDF determines the running of $\alphas$ and the value $\alphas(M_z)$ and
as before, the CTEQ6L PDF set was chosen. The ME calculations were
generated with fixed renormalisation and factorisation scales, taken to be
the geometric mean of the masses of the two heavy final-state particles.
The results of these choices remain evident in the ME-derived jet
distribution. In eqs.~(\ref{eq:py_evol}) and (\ref{eq:py_ISR_evol}) it was
shown that the \tsc{Pythia} shower algorithm instead picks scales related
to the the evolution variable when generating emissions. These differences
can be non-negligible, especially in the low-$\pT$ region, where $\alphas$
will become large and we expect a similar turnover to the \tsc{Pythia}
distribution. To account for these effects, additional weights are applied
to MadEvent distributions after generation, with the new scale taken to be 
the kinematic $\pT$ of the jet.

A correction for $\alphas$ is simple to achieve, weighting events by the
ratio of the new and old $\alphas$ values. With $M$ defined as the fixed
renormalisation/factorisation scale used in the ME calculations, we have a
weight
\begin{equation}
\frac{\alphas(\pT^2)}{\alphas(M^2)}
~.
\end{equation}
With ISR, correcting for differences in the factorisation scale is
more difficult. The distribution for real emission from the ME
calculation contains two PDF factors from the incoming partons, unlike
ISR generated from the shower algorithm. This difference makes it unclear
what correction should be applied, especially as in our ME calculation, we
do not know which incoming parton has branched. Taking an exclusive 
point of view, there should have been no emissions between $M^2$ and
$\pT^2$ on either side of the event, and so both should be reweighted,
overall giving a correction factor
\begin{equation}
\frac{x_1 f_1(x_1, \pT^2)}{x_1 f_1(x_1, M^2)}~
\frac{x_2 f_2(x_2, \pT^2)}{x_2 f_2(x_2, M^2)} ~.
\end{equation}
This may be a slight over-correction, however, and one would expect the
results with just the $\alphas$ factor and those with this additional PDF
factor to bracket the ``correct'' distribution. We examine this further in
Sec.~\ref{sec:top} for the case of top pair production.

\subsubsection{Graph topologies}
\label{sec:graphs}
\begin{figure}
\begin{center}
\begin{minipage}{5.2cm}
\begin{center}
\includegraphics[scale=0.6,trim=0 12 200 595,clip=true]{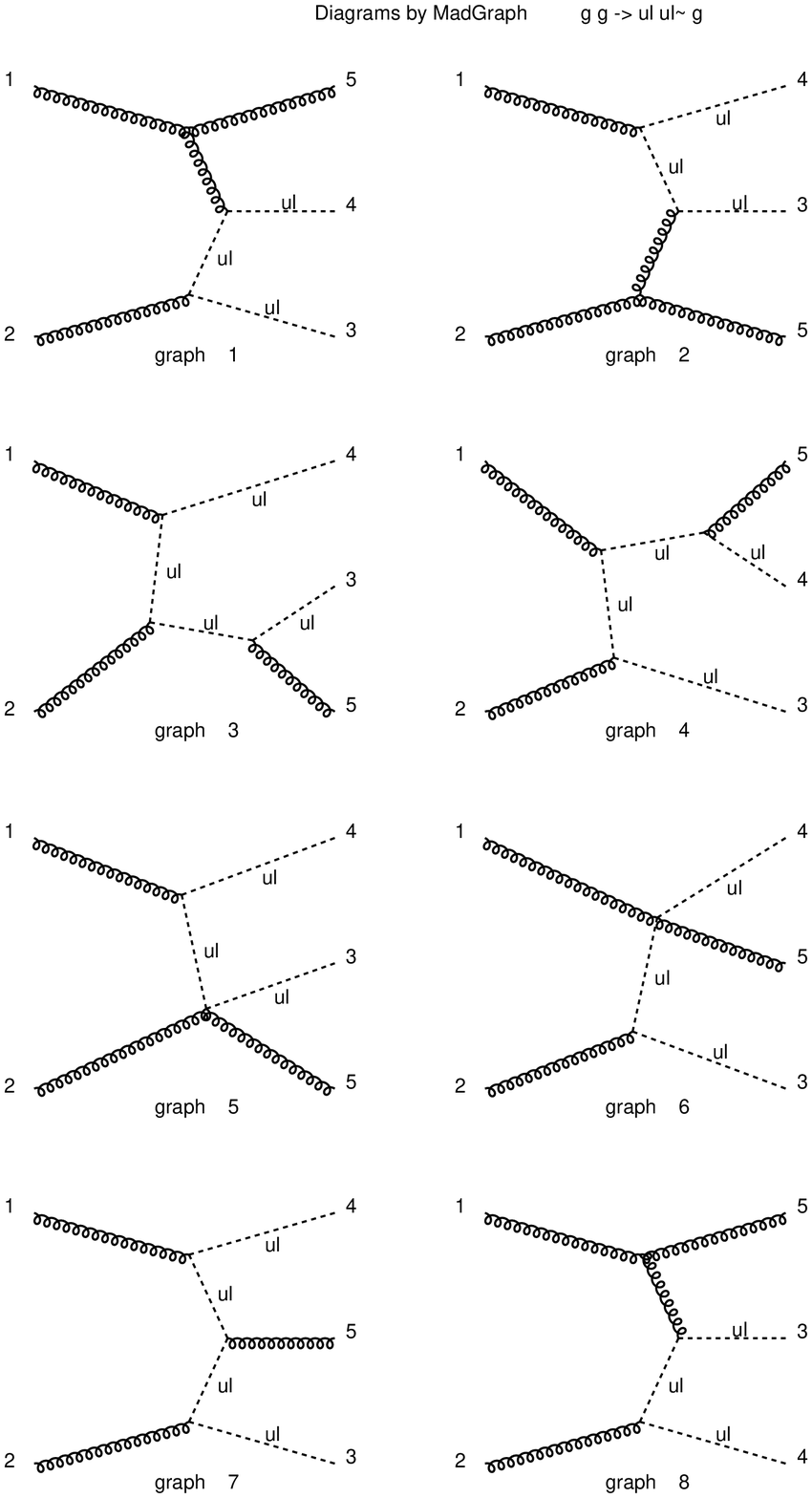}
\\\vspace{1mm}\hspace{-4mm}\scriptsize{(a)}
\end{center}
\end{minipage}
\begin{minipage}{5.2cm}
\begin{center}
\includegraphics[scale=0.6,trim=200 575 0 30,clip=true]{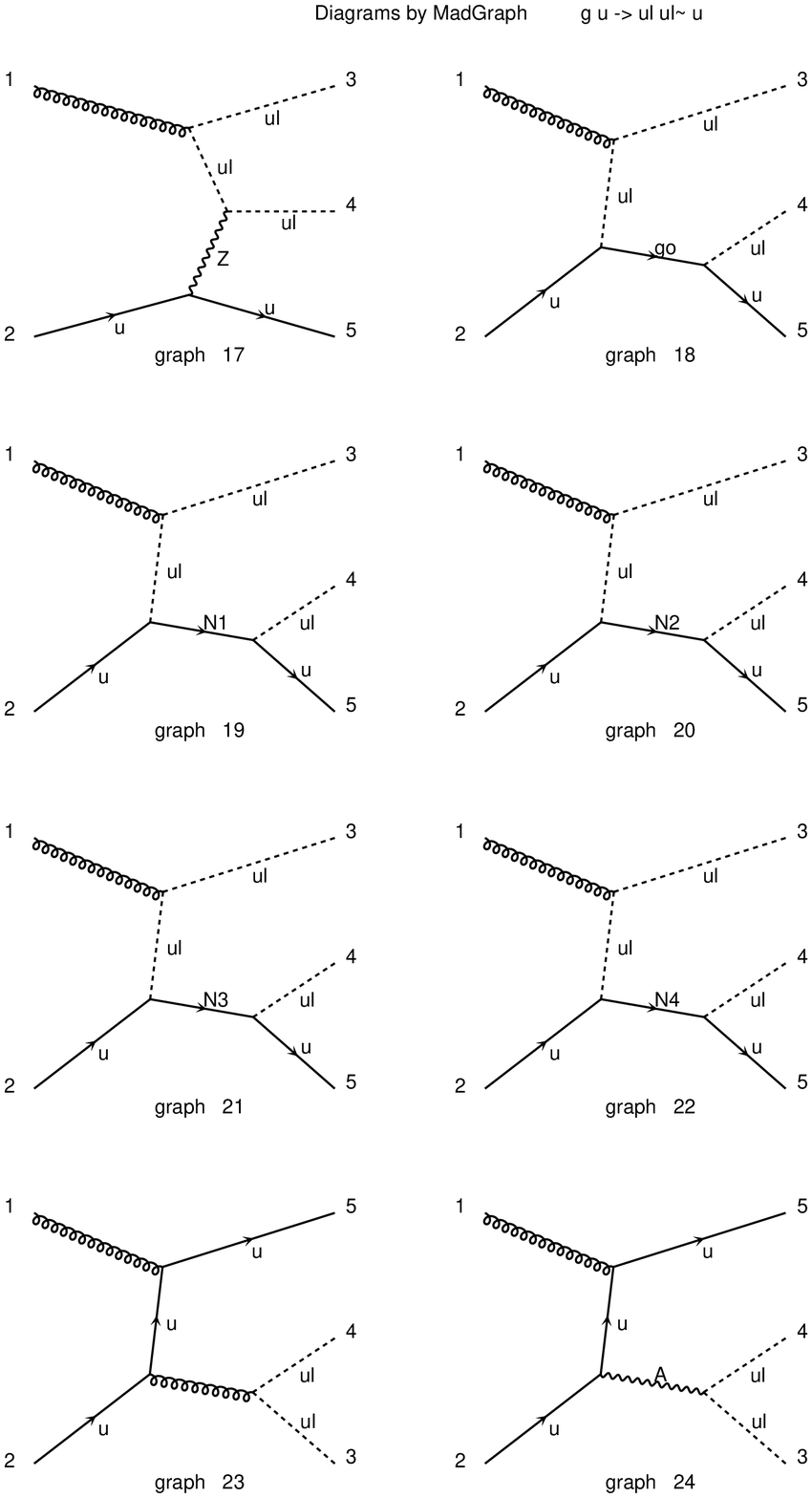}
\\\vspace{1mm}\hspace{7mm}\scriptsize{(b)}
\end{center}
\end{minipage}
\caption{Two possible graphs for $\squ_L$ squark pair production with an
additional jet which do not directly correspond to a $2 \to 2$ process and
a \tsc{Pythia} shower splitting
\label{fig:graph-squark}
}
\end{center}
\end{figure}

When moving to a final state containing an extra jet, MadEvent will
correctly generate all possible topologies, but when comparing against
\tsc{Pythia}, some of these graphs may not correspond to a shower history.
Two examples are given in Fig.~\ref{fig:graph-squark} for the case of up
squark pair production in the MSSM.

In (a), the two t-channel squark propagators cannot directly be reproduced
by a $2 \to 2$ process and a shower emission. In a traditional merging
approach, these topologies can always be associated to some $2 \to 2$ hard
process with the radiation associated to an incoming leg. As radiation off
a massive leg is suppressed, topologies of this kind are not expected to
dominate the overall behaviour of the real-emission matrix elements.

Instead, in (b), the squark and jet are created through the resonant decay of a
gluino. This resonant mechanism leads to events with a different topology
than we wish to consider; as \tsc{Pythia} works in the narrow width
approximation, where the gluino production and decay are described as part
of the separate $2 \to 2$ process $\u \g \to \tilde{\u} \tilde{\g}$, 
it would be doublecounting to include the same graph as a $2 \to 3$ process. 

To remove double counting of this type, we use the method outlined in
\cite{Alwall:2008qv}; the idea here is to assign propagators to events
statistically, based on the relative size of the relevant squared
amplitudes. Such events then explicitly contain this propagator
information in the LHEF output, meaning they can be easily vetoed when
subsequently being processed. As a simple check that the
results are reasonable, many processes were also generated with these
resonant graphs manually removed, e.g.\ as in \cite{Plehn:2005cq}. Although
this procedure is not fully gauge invariant, as the widths of the resonances are
small, we expect the results to be in line with the MadEvent veto method.

\subsection{Standard Model processes}
\label{sec:top}

We begin by comparing the jet emission probability in top pair production of
POWHEG, MadEvent and the first \tsc{Pythia} shower emission (ISR only).
In all studies that follow, events are generated at LHC energies ($\p\p$,
$\sqrt{s} = 14\TeV$) and, additionally, when the damping ansatz is used, the
scale $M^2$ of eq.~(\ref{eq:pTdamp}) is set to be the factorisation scale
unless stated otherwise. For top production, we restrict both \tsc{Pythia}
and MadEvent to QCD production only, as this is the dominant contribution
to the cross section, and also allows a consistent comparison to POWHEG. As
noted previously, the default \tsc{Pythia} ISR shower will begin evolution
at the kinematical limit, given that a top cannot be produced in the
shower. The POWHEG to MadEvent comparison allows us to check the validity
of the MadEvent approximation method outlined in Sec.~\ref{sec:me-method}.
The comparison to \tsc{Pythia} then lets us examine the effectiveness of
the ansatz described in Sec.~\ref{sec:py-correction}.

\begin{figure}
\centerline{
\includegraphics[angle=270,scale=0.63]{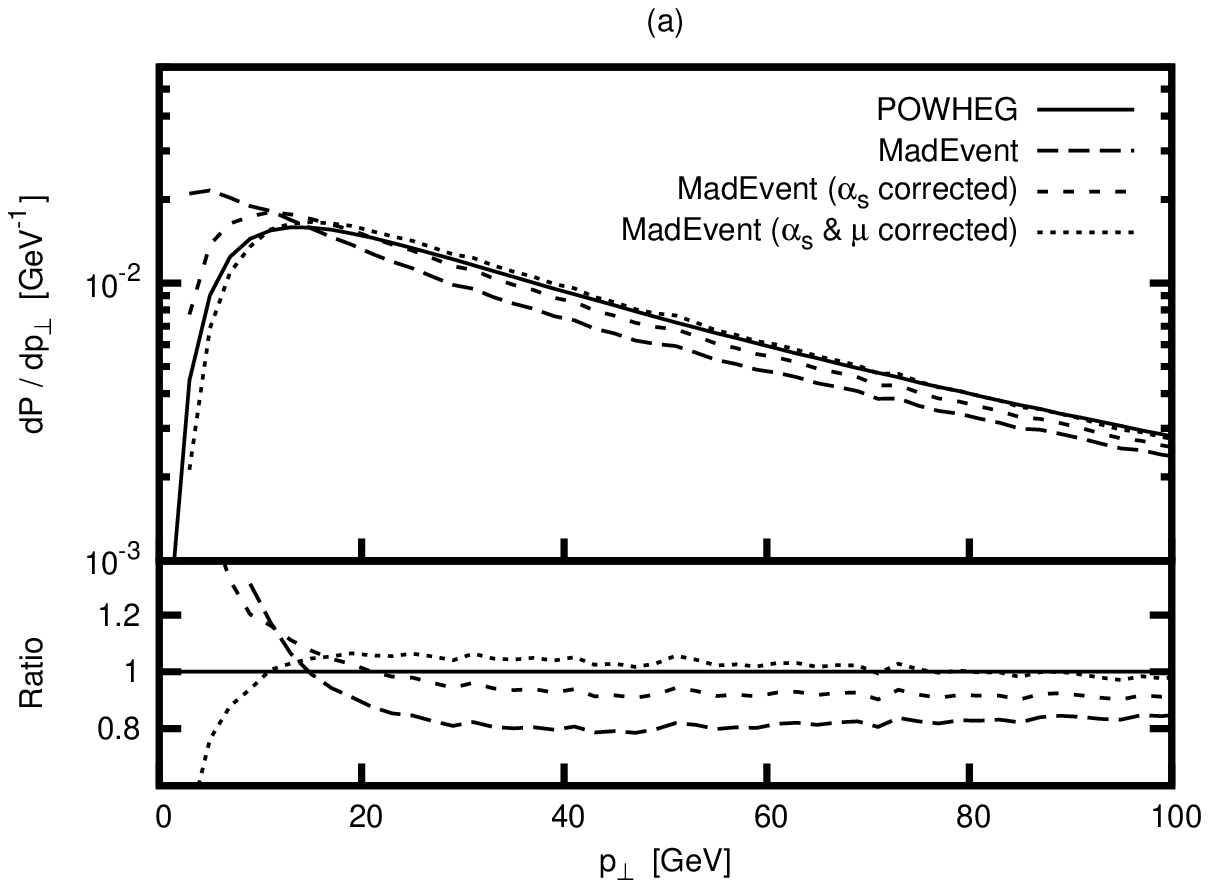}
\includegraphics[angle=270,scale=0.63]{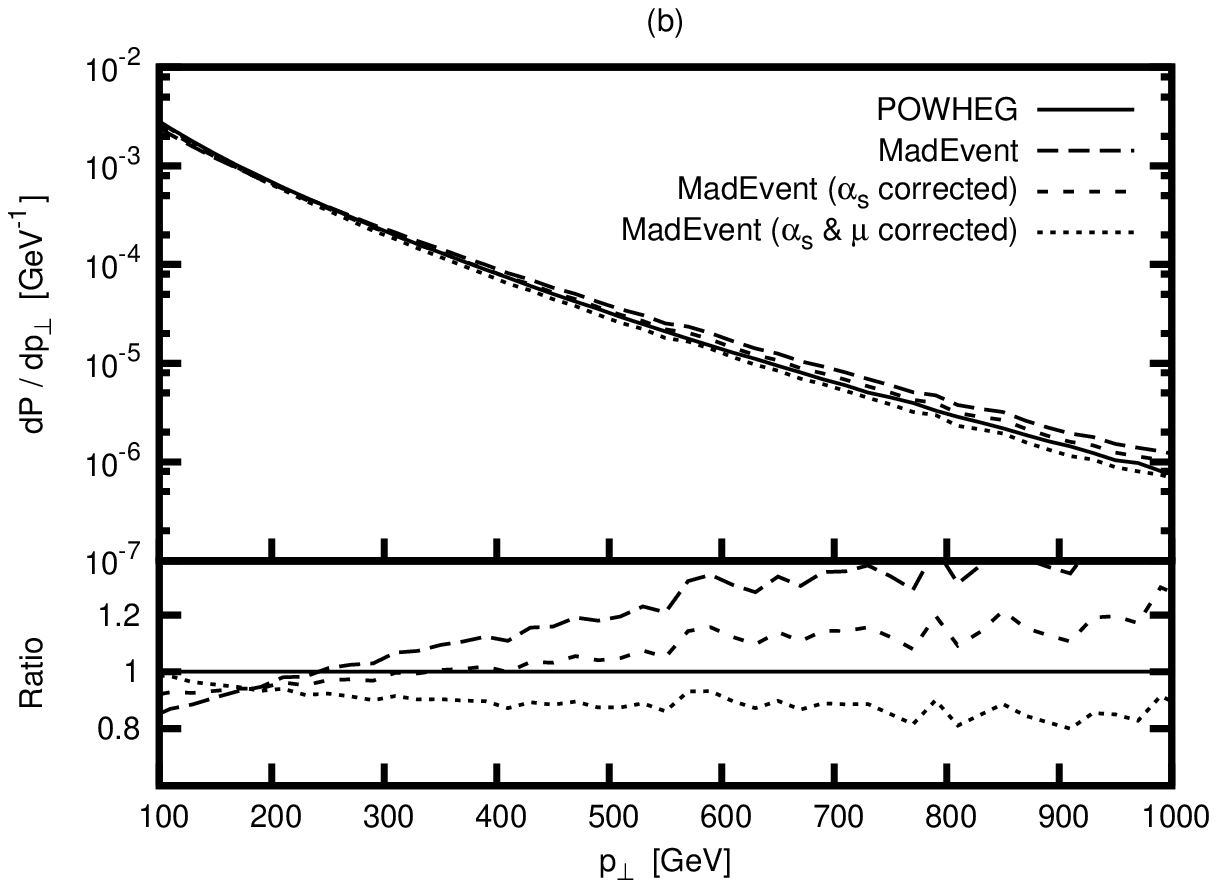}
}
\caption{First emission probability as a function of $\pT$ for top pair
production in POWHEG and MadEvent. Ratio plots are normalised to the POWHEG
result
\label{fig:tt-ph-me-ptjet}
}
\end{figure}

The probability distributions for jet emission in top pair production are
expected to be roughly in agreement over the entire $\pT$ range for POWHEG
and MadEvent, as long as the kinematical differences that come from the full
NLO prefactor of eq.~(\ref{eqn:POWHEG-he}) are small. This is shown in
Fig.~\ref{fig:tt-ph-me-ptjet}, separately for low and high $\pT$ regions,
for three different sets of the MadEvent results. All three have the Sudakov
correction applied, but one set additionally has the $\alphas$ scale
correction (``$\alphas$ corrected''), and the other both the $\alphas$ and
factorisation scale corrections (``$\alphas$ \& $\mu$ corrected'').
These will all meet when the $\pT$ scale matches that
of the fixed factorisation/renormalisation scale used in the MadEvent
generation, but begin to diverge as the $\pT$ rises and falls away from
this value.  In the low-$\pT$ region, the Sudakovs make the distributions
turn over, while in the high $\pT$ tail, there is good agreement between
the distributions of POWHEG and the approximation from MadEvent; in the
ratio plot the POWHEG curve sits between the two different
corrected MadEvent curves, but overall closer to the $\alphas$ \&
$\mu$-corrected one.

\begin{figure}
\centerline{
\includegraphics[angle=270,scale=0.63]{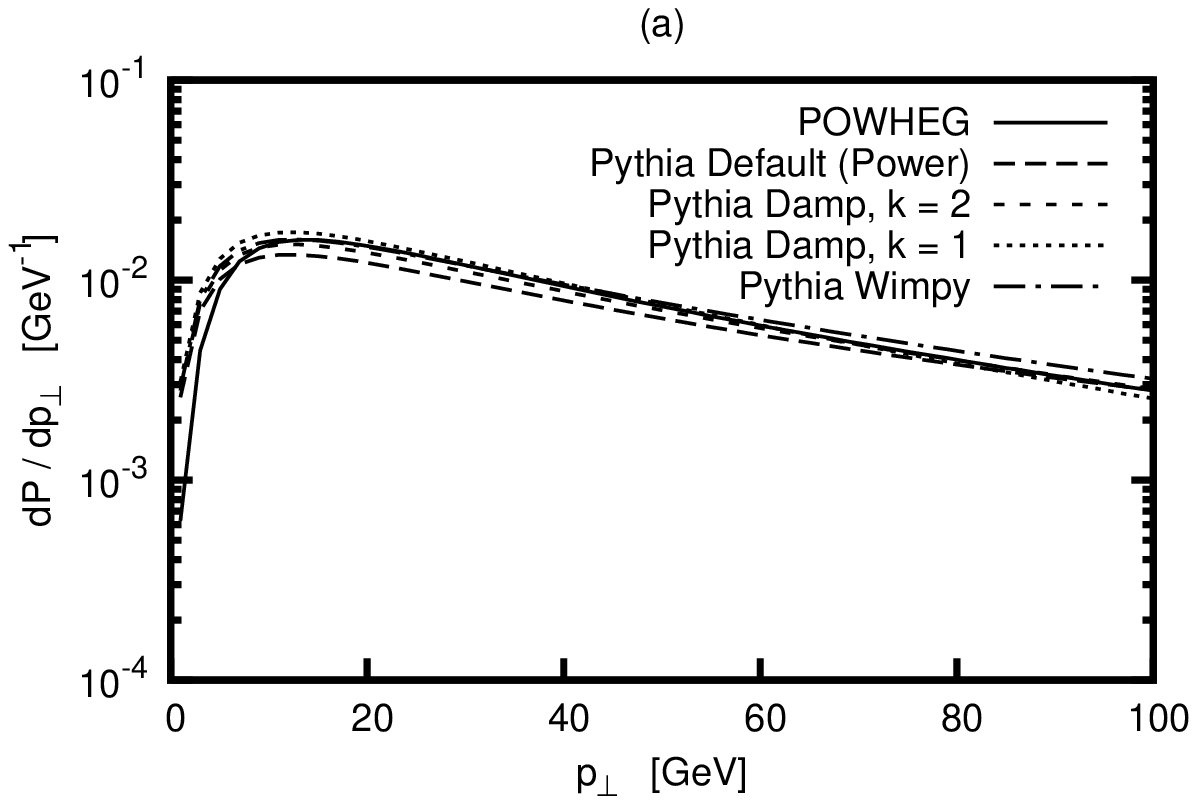}
\includegraphics[angle=270,scale=0.63]{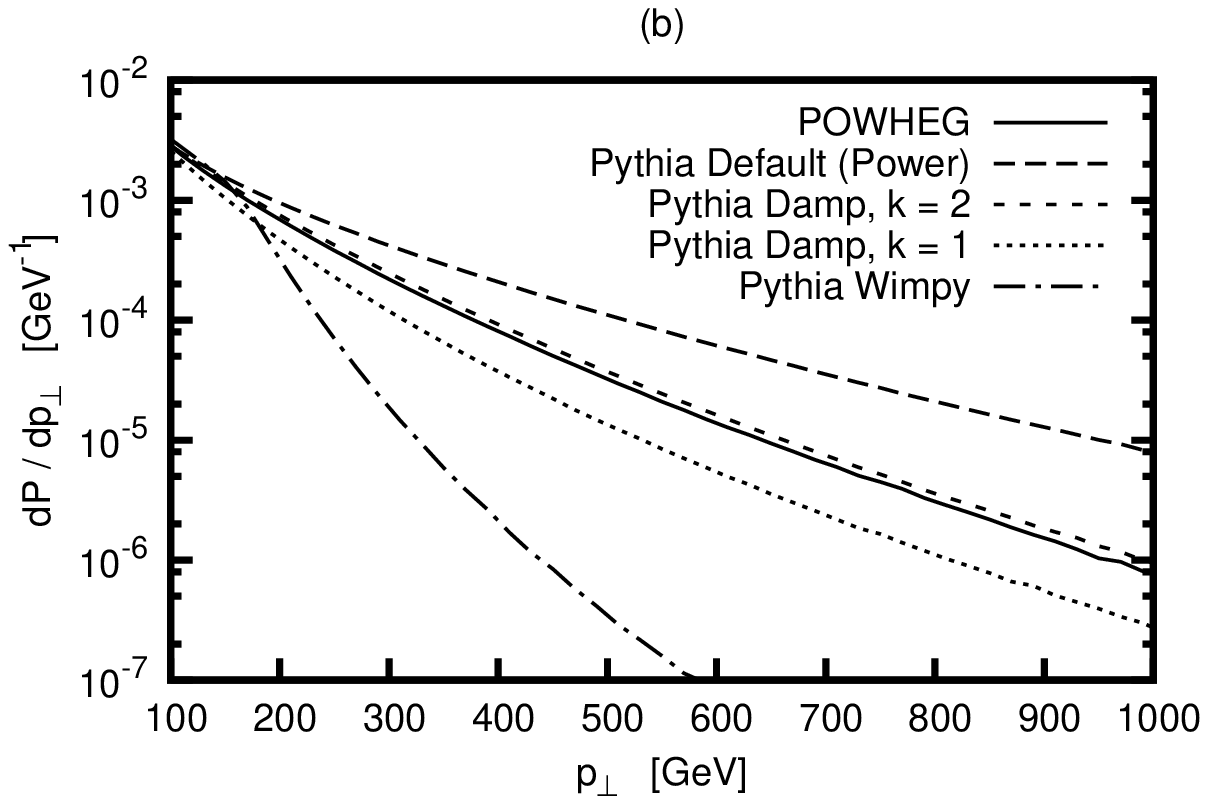}
}
\caption{First emission probability as a function of $\pT$ for top pair
production in POWHEG and \textsc{Pythia}, split into (a) the low-$\pT$
region where the Sudakov makes the distributions turn over and (b) the
high-$\pT$ tail
\label{fig:tt-ph-py-ptjet}
}
\end{figure}

We now move on to compare POWHEG against the \tsc{Pythia} shower and the
ansatz of Sec.~\ref{sec:py-correction}. The expectation is that the
default \tsc{Pythia} power shower will not fall off quickly enough and therefore
overestimate the emission probability in the high $\pT$ tail. In
Fig.~\ref{fig:tt-ph-py-ptjet} we show, again separately for low and high
$\pT$ regions, the POWHEG results against four variations of the
\tsc{Pythia} shower (wimpy, power, damped with $k = 1$ and damped with $k =
2$). Again, there is good agreement in the low-$\pT$ region. In the
high-$\pT$ tail, the wimpy shower sits clearly below the POWHEG result,
while the \textsc{Pythia} power shower does indeed overestimate the jet
emission probability. The damping procedure, particularly with $k = 2$,
then brings the \textsc{Pythia} distribution into closer agreement with
POWHEG.

\begin{figure}
\centerline{
\includegraphics[angle=270,scale=0.63]{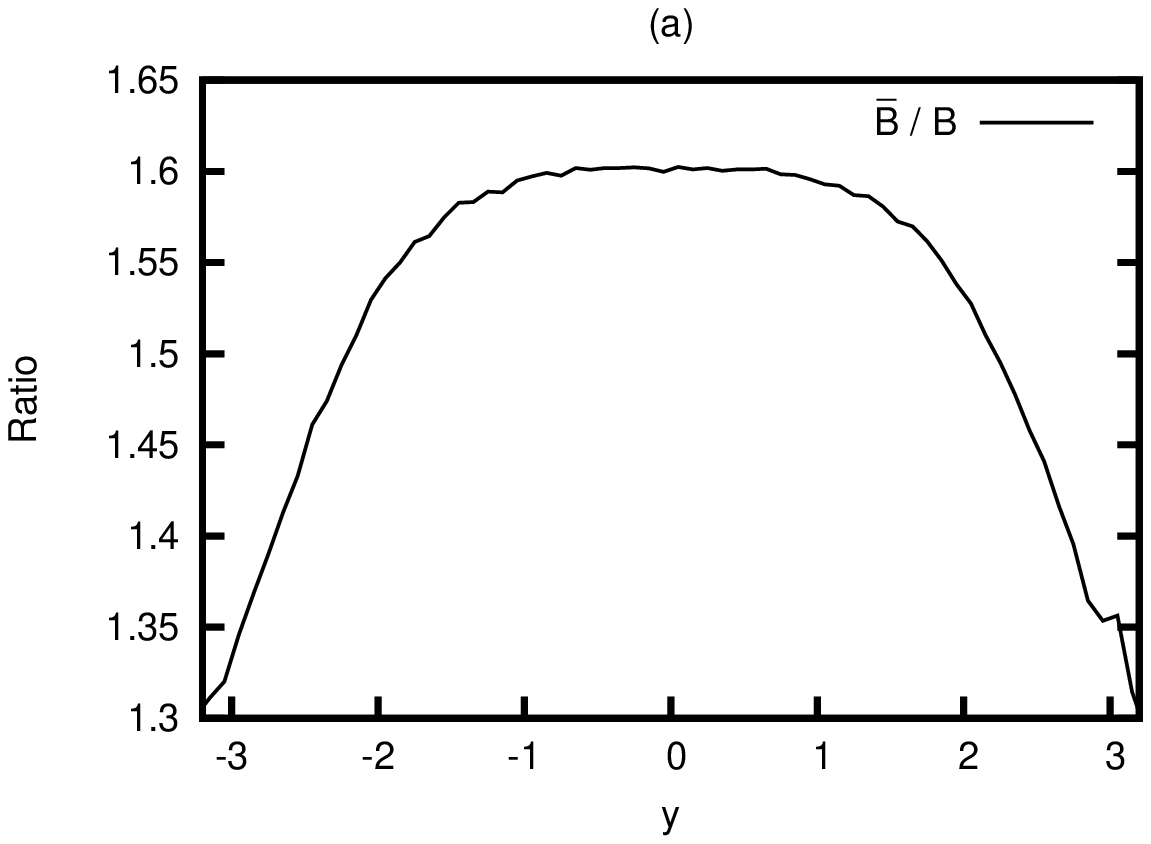}
\includegraphics[angle=270,scale=0.63]{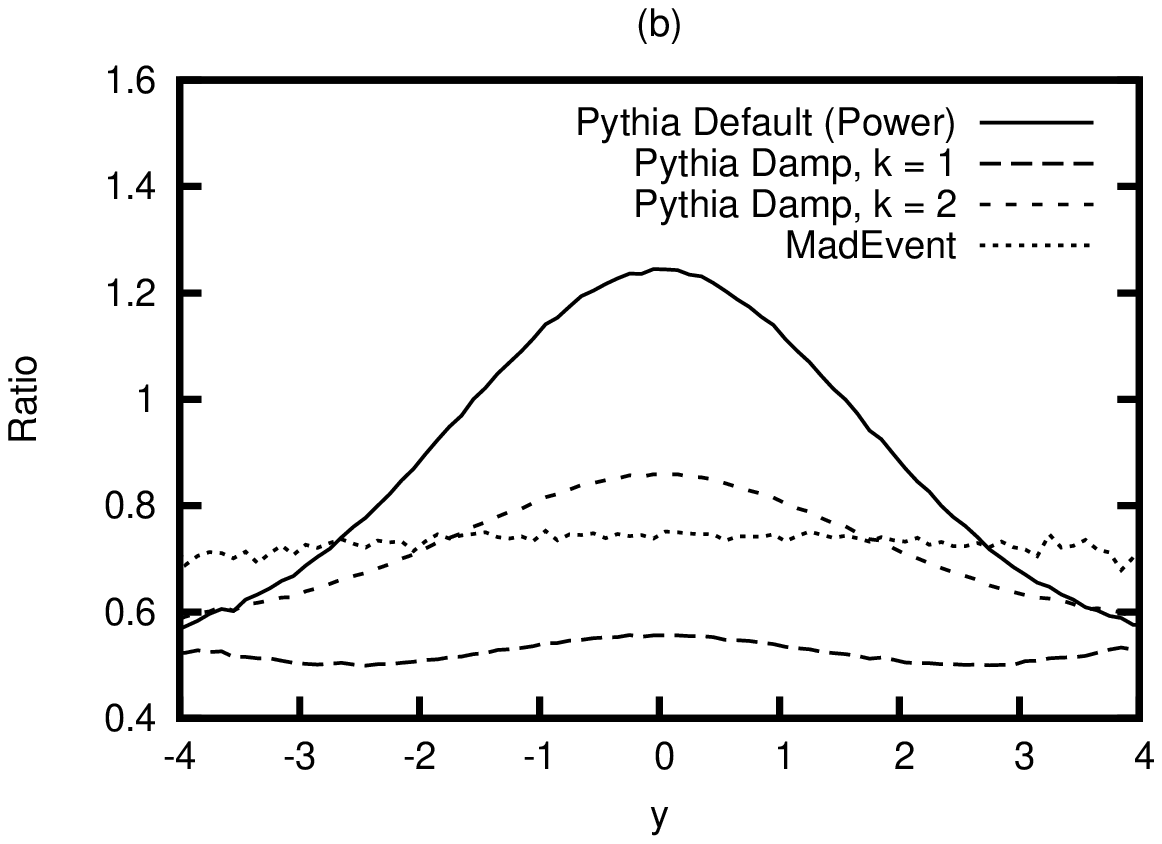}
}
\caption{(a) Ratio of top pair rapidity between POWHEG with the full NLO
prefactor, $\bar{B}$, of eq.~(\ref{eqn:POWHEG-he}) and $B$ only. 
Note the suppressed zero on the $y$ axis. (b)
Jet rapidity distributions of \textsc{Pythia} and MadEvent for ${\pT}_{jet}
> 100 \mrm{GeV}$, normalised to the POWHEG result
\label{fig:ttbar-rap}
}
\end{figure}

It is interesting to study the kinematical differences that come from
the full NLO prefactor in POWHEG by considering the top pair rapidity both
with and without such a prefactor. When full NLO corrections are included,
we not only expect an overall change in cross section, but also that there
may be kinematic differences in the distributions. In
Fig.~\ref{fig:ttbar-rap}a, the ratio of two POWHEG pair rapidity
distributions is shown. The first is from the default POWHEG-hvq generator,
with a full NLO prefactor, while the second is generated by replacing the
$\bar{B}$ term of eq.~(\ref{eqn:POWHEG-he}) with just the Born contribution
$B$. The figure shows both features; an overall shift in
cross section (a $k$-factor of around 1.5) and a small shift in pair
rapidity to more central regions.

We finally compare the jet rapidity distributions from POWHEG, MadEvent and
\tsc{Pythia}. In \tsc{Pythia}, the rapidity of the extra jet is taken
immediately after the first shower emission (if present). For MadEvent,
however, we can not apply the Sudakov correction in this case, and the
direct output of the generator is taken. We restrict ourselves to
${\pT}_{jet} > 100 \mrm{GeV}$, such that the effect of the Sudakov and
other scale corrections will be small. Here it is harder to know what 
to expect. For the damped \tsc{Pythia} shower, it is clear that by only 
reducing the high-$\pT$ tail relative to the power shower,
events will be taken out of the central region. 
These distributions are shown in Fig.~\ref{fig:ttbar-rap}b, normalised
to the POWHEG result. MadEvent and POWHEG are in good
agreement, although with an overall shift in cross section. The default
\tsc{Pythia} shower is strongly peaked in the central region, but
comes into better agreement when damped.

\begin{figure}
\centering
\includegraphics[angle=270,scale=0.63]{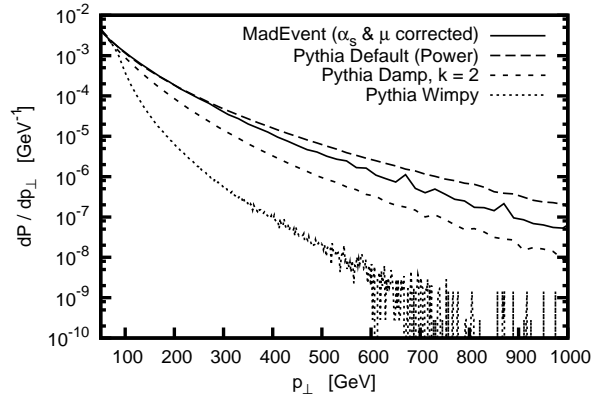}
\caption{First emission probability as a function of $\pT$ for Z pair
production. Results show \textsc{Pythia} compared to the
approximate MadEvent ($\alphas$ \& $\mu$ corrected) prescription
\label{fig:Zboson-ptjet}
}
\end{figure}

We now study processes where there is not (yet) a ``correct'' POWHEG result
to compare with, but instead rely on the MadEvent approximation.
If the ansatz of Sec.~\ref{sec:py-correction} is correct,
the damping of the \tsc{Pythia} shower is not expected to help improve the
$\pT$ distribution in Z and W pair production in a significant way. Here,
there are no coloured final-state particles to use as a guide in setting the
$M^2$ scale of eq.~(\ref{eq:pTdamp}), so we continue to use the
factorisation scale when generating the damped results.
Again, in this case, the \tsc{Pythia} default is to start ISR shower at the
kinematical limit. Fig.~\ref{fig:Zboson-ptjet} shows the results for
Z pair production for the default, damped ($k = 2$) and wimpy shower
against MadEvent ($\alphas$ \& $\mu$ corrected). The results show that the
default shower, although giving a slightly too hard $\pT$ tail, still does
a reasonable job of reproducing the MadEvent curve without additional damping.

\begin{figure}
\centerline{
\includegraphics[angle=270,scale=0.63]{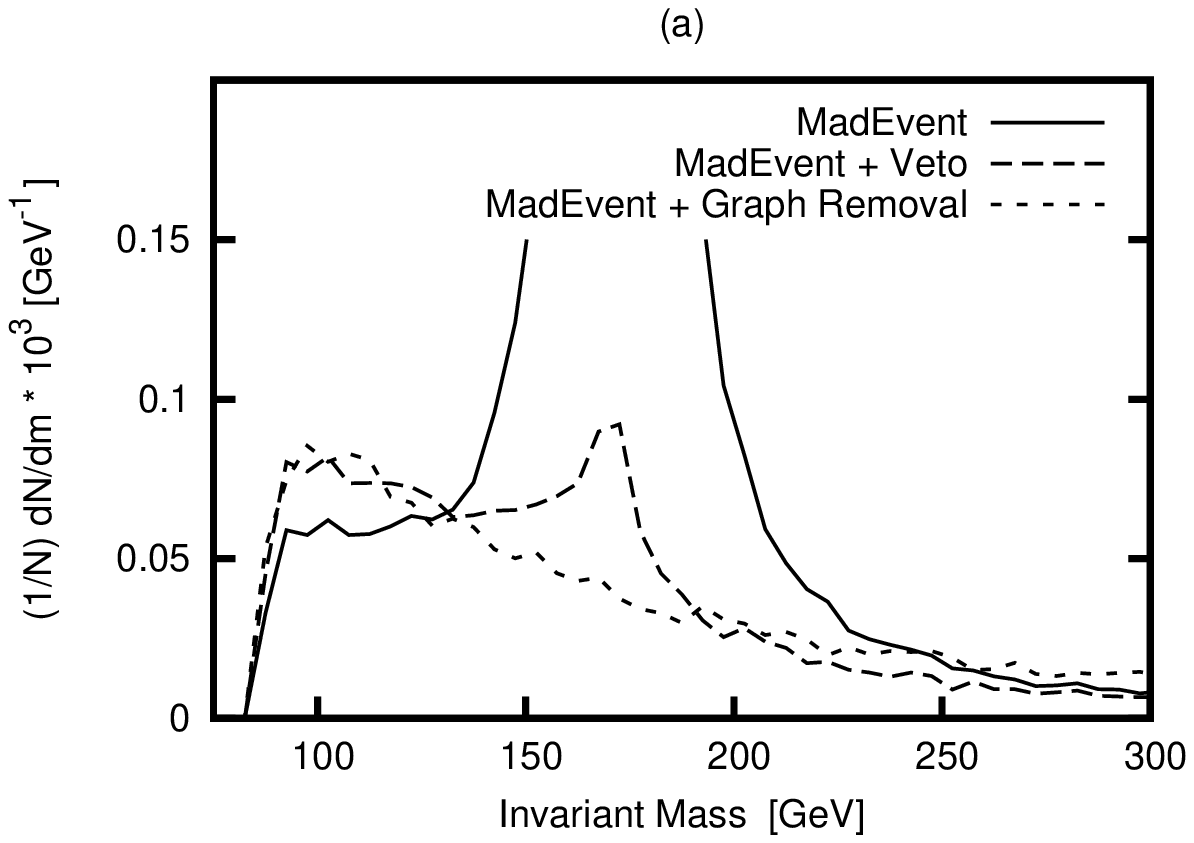}
\includegraphics[angle=270,scale=0.63]{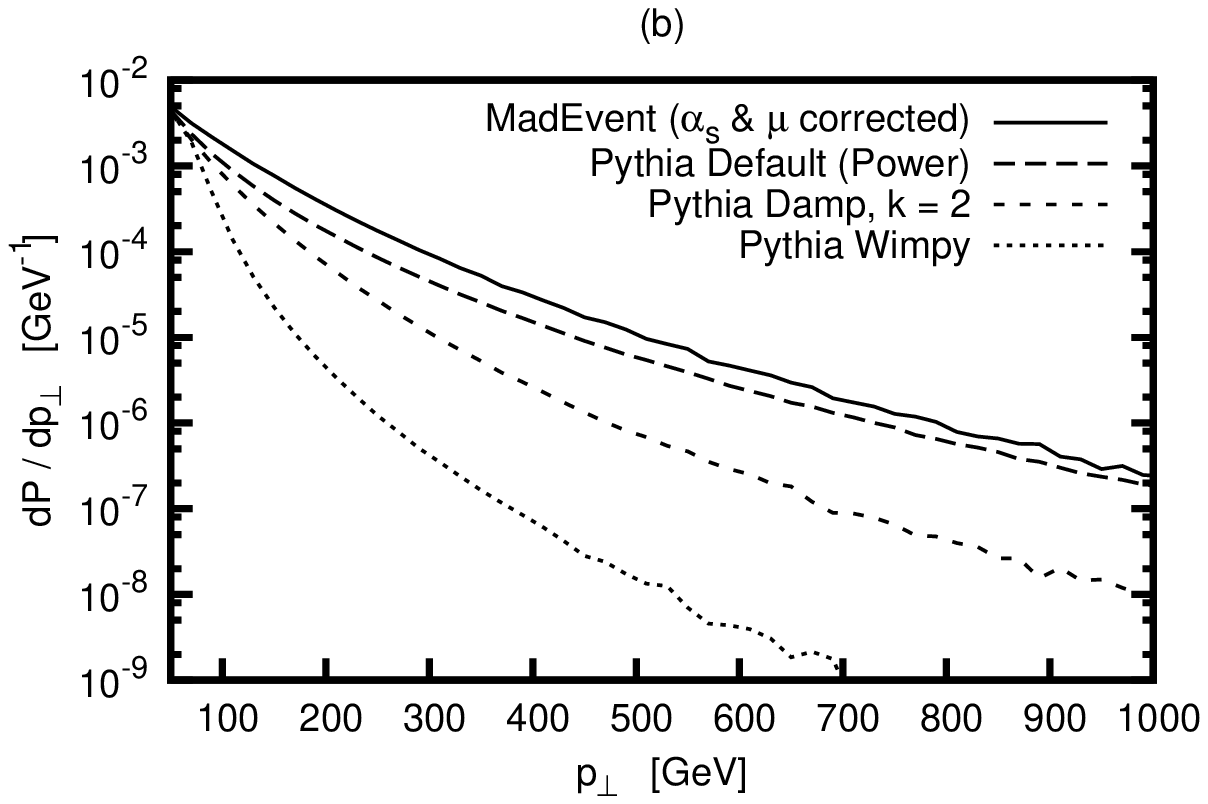}
}
\caption{$\W$ pair production. (a) Shows the invariant mass of the jet and
$\W$ boson when the jet is a bottom quark for three different MadEvent runs
(see text). The large resonant peak in the ``MadEvent'' sample has been cut
off above 0.15. (b) Shows the first emission probability as a function of
$\pT$
\label{fig:Wboson-ptjet}
}
\end{figure}

The case of W pair production is useful as a check on the MadEvent
resonance veto method (see Sec.~\ref{sec:graphs}). When bottom quarks
are allowed in both the incoming beams and in the definition of the extra
jet, there is now a resonant contribution coming from $\t \to \b\W$. In
Fig.~\ref{fig:Wboson-ptjet}a, the invariant mass of the jet and the
matching $\W$ are shown when the jet is a bottom quark. A clear peak is
visible when those events with a resonance are not vetoed (``MadEvent''),
while this peak is almost completely removed when these events are vetoed 
(``MadEvent + Veto''); there is only a small remaining trace of
it from e.g.\ interference terms. As an additional check, also plotted is
the result when resonant graphs are explicitly removed from the MadEvent
generation (``MadEvent + Graph Removal''). Fig.~\ref{fig:Wboson-ptjet}b,
shows the results for the first emission probability in the $\pT$ tail for
the MadEvent vetoed sample. As a final check, the radiation pattern was
compared to a sample not including bottom quarks, and found to contain no
large differences.

\subsection{MSSM studies}

To further study the production of heavy final states, MSSM
processes were chosen split into three groups: coloured final states,
non-coloured final states, and coloured/non-coloured final states. All
events were generated with the SPS1a \cite{Allanach:2002nj} set of
parameters (the relevant final-state masses used are shown in
Tab.~\ref{tab:MSSMmasses}). Here, also the lowest order $2 \to 2$ processes
were generated using MadEvent, and fed into \textsc{Pythia} for
showering, again with the factorisation and renormalisation scales fixed at
the geometric mean of the masses of the two heavy final-state particles.
Note that this is slightly different from the default \tsc{Pythia}
internal $2 \to 2$ process scale choices and that this choice enters into
the $M^2$ scale selection of the damping ansatz, eq.~\ref{eq:pTdamp}.
 
\begin{table}
\centering
\begin{tabular}{|c|c|}
\hline
\bf{Particle} & \bf{Mass (GeV)} \\ \hline
${\squ}_L$      & 561.1 \\ 
${\sg}$         & 607.7 \\ 
${\schi}_1^0$   & 96.7  \\ 
${\schi}_1^\pm$ & 181.7 \\ \hline
\end{tabular}
\caption{Subset of MSSM SPS1a masses used in process generation
\label{tab:MSSMmasses}}
\end{table}

\subsubsection{Coloured final states}

\begin{figure}
\centering 
\includegraphics[angle=270,scale=0.63]{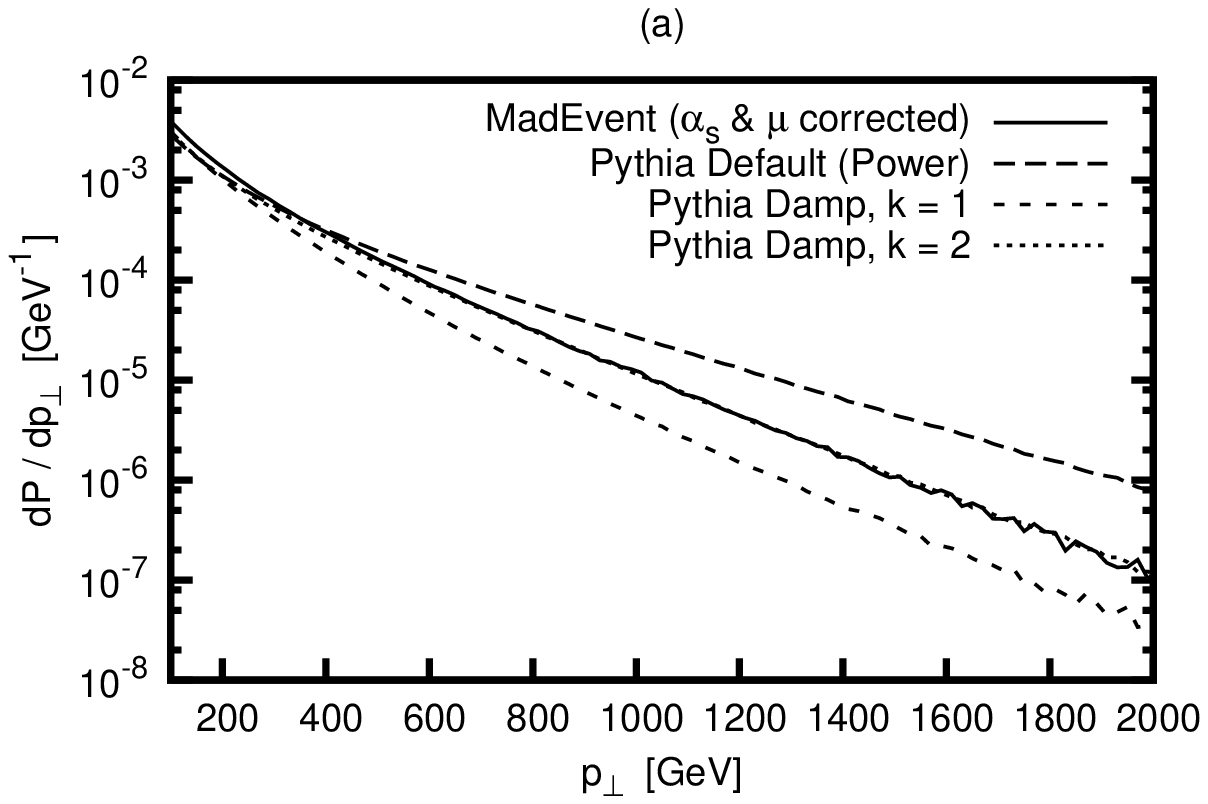}
\centerline{
\includegraphics[angle=270,scale=0.63]{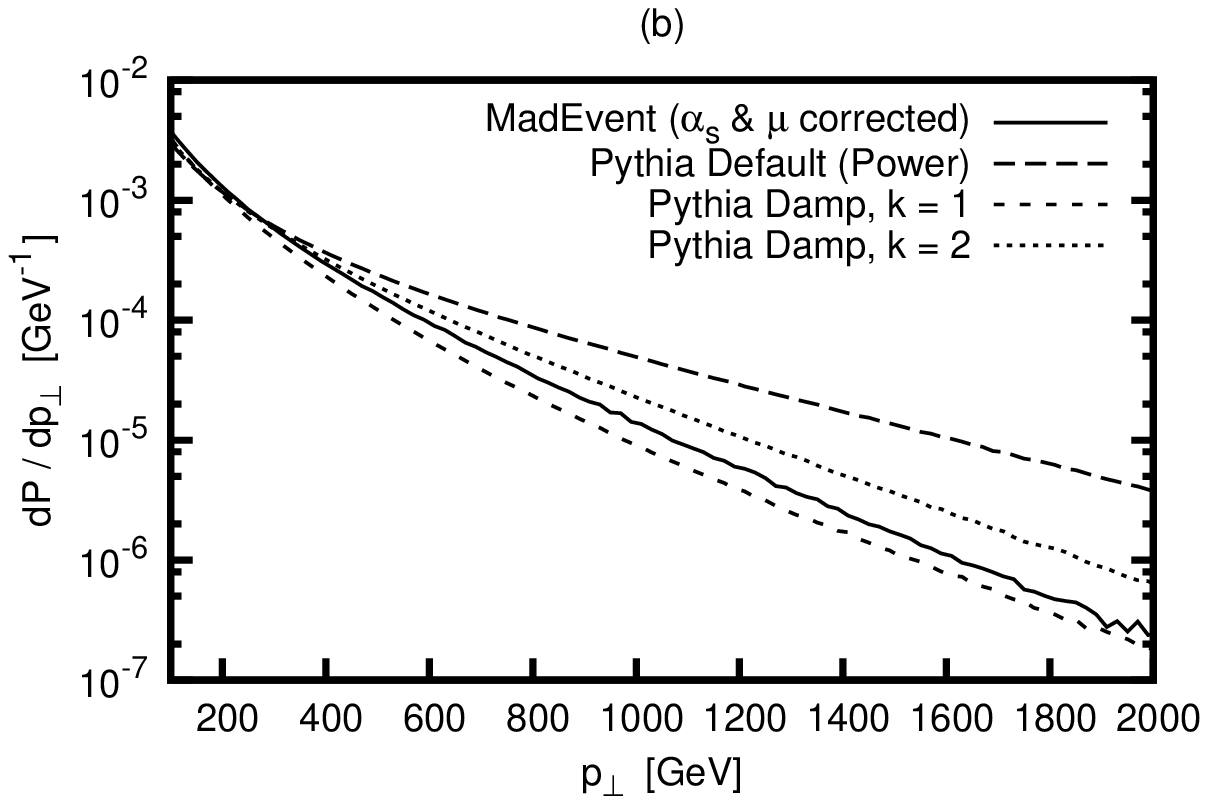}
\includegraphics[angle=270,scale=0.63]{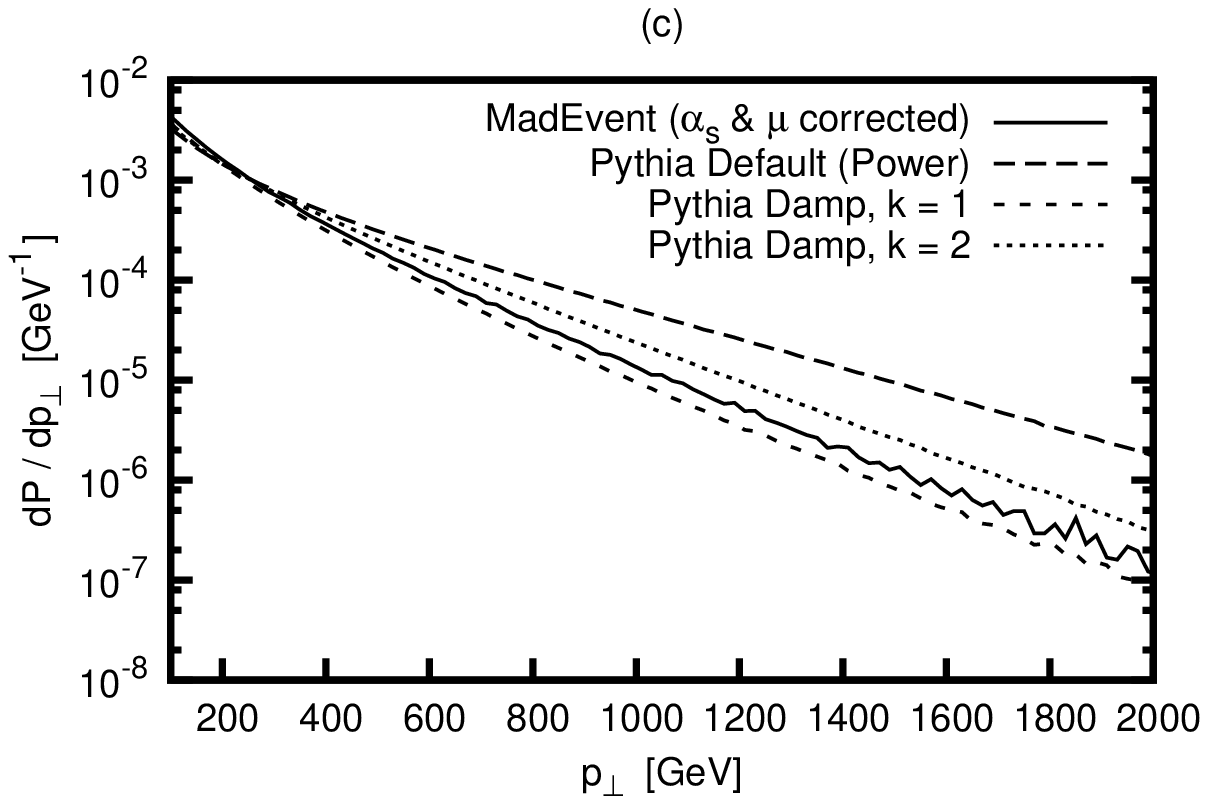}
}
\caption{First emission probability as a function of $\pT$ for (a)
${\squ}_L \, \bar{\squ}_L$, (b) ${\squ}_L \, \sg$ and (c) $\sg \, \sg$
production. In (a) the $k = 2$ and MadEvent curves lie on top of each other
\label{fig:MSSM-pT-1}
}
\end{figure}

For the fully coloured final states, all events were generated with QCD
only; as with top pair production, this is where the dominant contribution
to the cross section lies. As before, the expectation is that the
coloured final states will benefit from a damping of the high-$\pT$ tail.
Figure \ref{fig:MSSM-pT-1} shows the tail of the $\pT$ distributions, all
compared to the MadEvent $\alphas$ \& $\mu$ corrected data, for
(a) ${\squ}_L \, \bar{\squ}_L$, (b) ${\squ}_L \, \sg$ and (c) $\sg \, \sg$.
Similarly to top pair production, the power shower overestimates
the high-$\pT$ tail, while the damping ansatz brings the curves into better
agreement. For ${\squ}_L \, \bar{\squ}_L$ production, $k = 2$
leads to the best agreement, while for ${\squ}_L \, \sg$ and
$\sg \, \sg$ production, the MadEvent curve lies between the damped $k = 1$
and $k = 2$ curves.

\subsubsection{Non-coloured final states}

We move on to study the non-coloured final states, ${\schi}_1^0 \,
{\schi}_1^0$ and ${\schi}_1^+ \, {\schi}_1^-$, where we expect the damping
ansatz not to improve the $\pT$ tail of the parton shower. Here, for both
processes, at lowest order, there are large resonant $\H^0$/$\A^0$
contributions. These are relatively long-lived intermediate states which
we expect to follow the rules for $2 \to 1$ singlet production, where
the shower already does a good job in covering the entire phase
space. For both processes then, the
MadEvent veto scheme is used to remove events of this type.

\begin{figure}
\centerline{
\includegraphics[angle=270,scale=0.63]{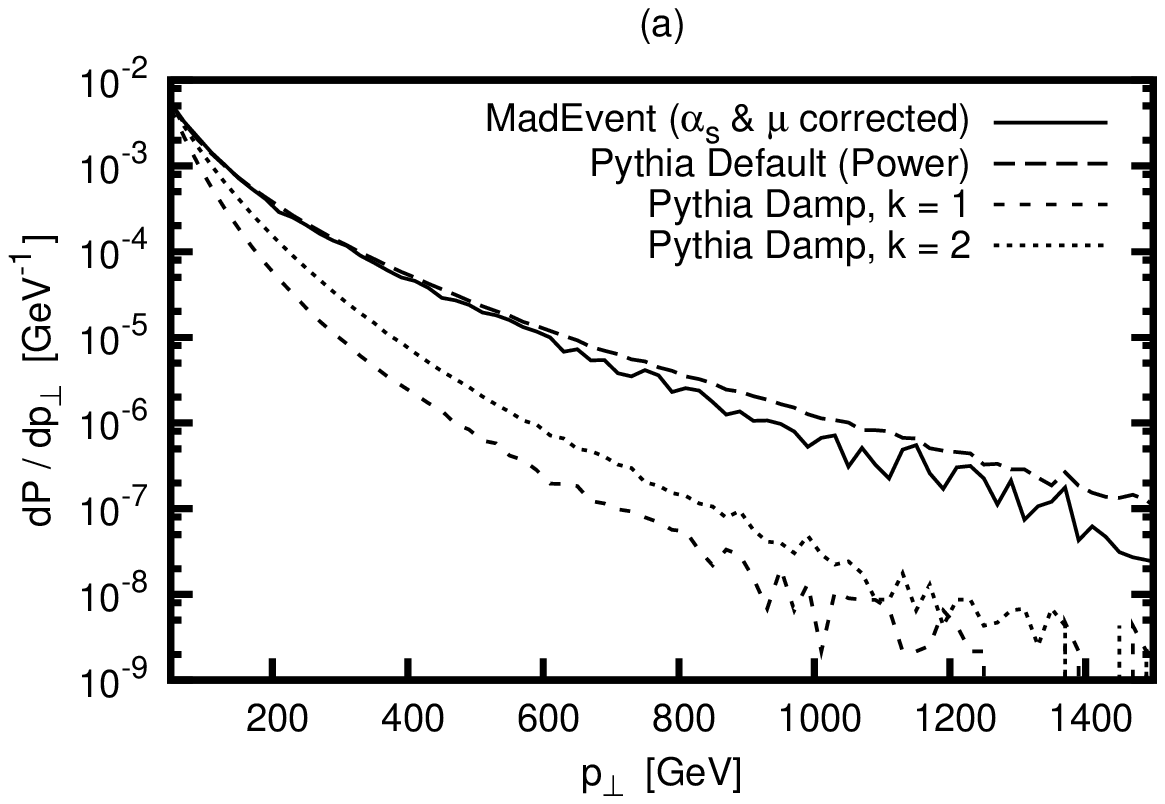}
\includegraphics[angle=270,scale=0.63]{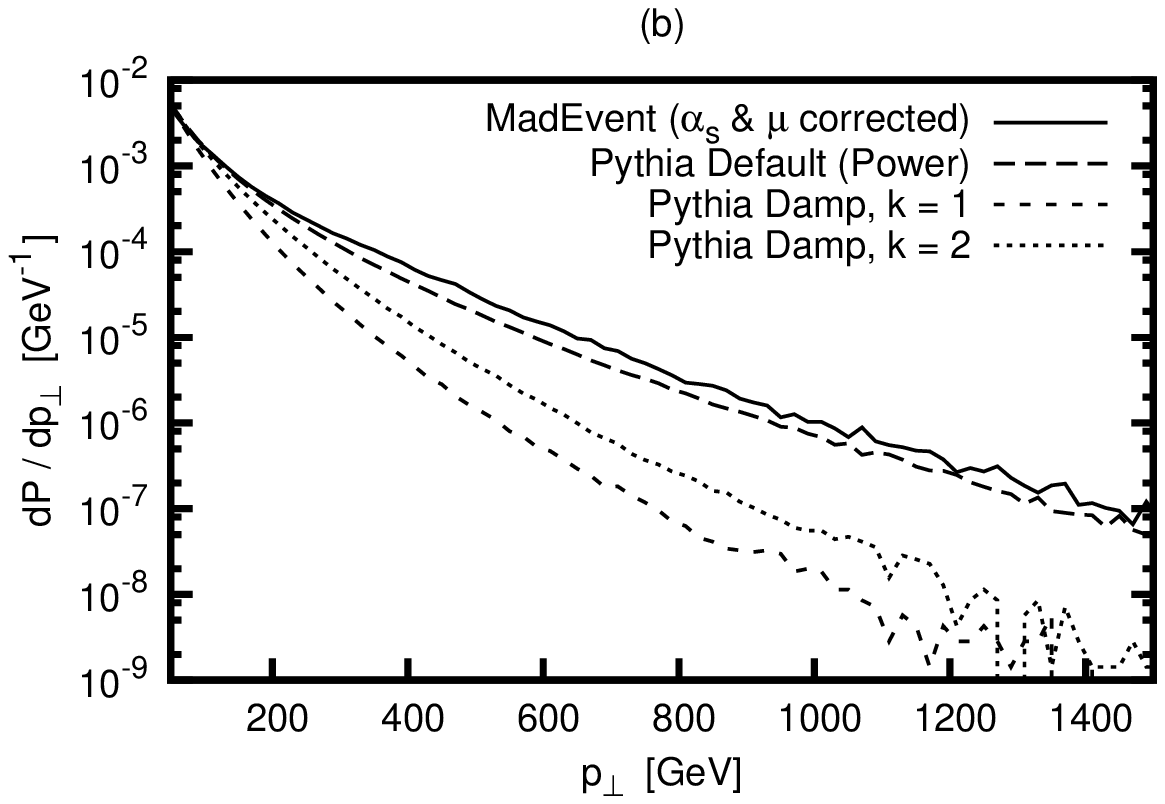}
}
\caption{First emission probability as a function of $\pT$ for (a)
${\schi}_1^0 \, {\schi}_1^0$ and (b) ${\schi}_1^+ \, {\schi}_1^-$
production
\label{fig:MSSM-pT-2}
}
\end{figure}

The results are shown in Fig.~\ref{fig:MSSM-pT-2} for
(a) ${\schi}_1^0 \, {\schi}_1^0$ and (b) ${\schi}_1^+ \, {\schi}_1^-$
production. The results follow the pattern for $\Z$/$\W^{\pm}$ pair
production; in both cases, the default power shower does a reasonable job
in the high-$\pT$ tail.

\subsubsection{Coloured/non-coloured final states}
\label{sec:mixed}
\begin{figure}
\centering
\includegraphics[angle=270,scale=0.63]{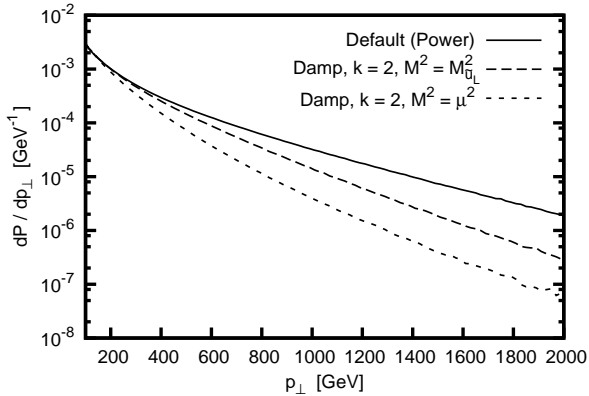}
\caption{\tsc{Pythia} first emission probability for the damped shower
($k = 2$) in ${\squ}_L \, {\schi}_1^0$ production, where $M^2$
is set either to the factorisation scale or to the mass of the heavy outgoing
squark. The default power shower curve is shown for comparison
\label{fig:MSSM-pT-scales}
}
\end{figure}

Finally, we study mixed coloured/non-coloured final states. Here, we would
expect some measure of damping to improve the shower description due to the
presence of colour in the final state. As discussed in
Sec.~\ref{sec:py-correction}, one issue here is the meaning of the
$M^2$ scale of eq.~(\ref{eq:pTdamp}). With the mixed final state, the
difference between the factorisation scale and e.g.\ just the mass of the
final-state coloured object is large enough to give noticeable differences
in the damped shower tail. As an example, Fig.~\ref{fig:MSSM-pT-scales}
shows the difference in the damped shower ($k = 2$) in
${\squ}_L \, {\schi}_1^0$ production, when $M^2$ is set to the
factorisation scale and when it is instead set to the mass of the outgoing
squark. In the remaining results of this section, we retain the choice of
setting $M^2$ to the mass of the outgoing coloured state. Also note that
all of the processes studied in this section have strong resonant
contributions, where a squark decays into a neutralino/chargino and a jet,
making it difficult to generate large statistics when using the MadEvent
veto scheme.

\begin{figure}
\centerline{
\includegraphics[angle=270,scale=0.63]{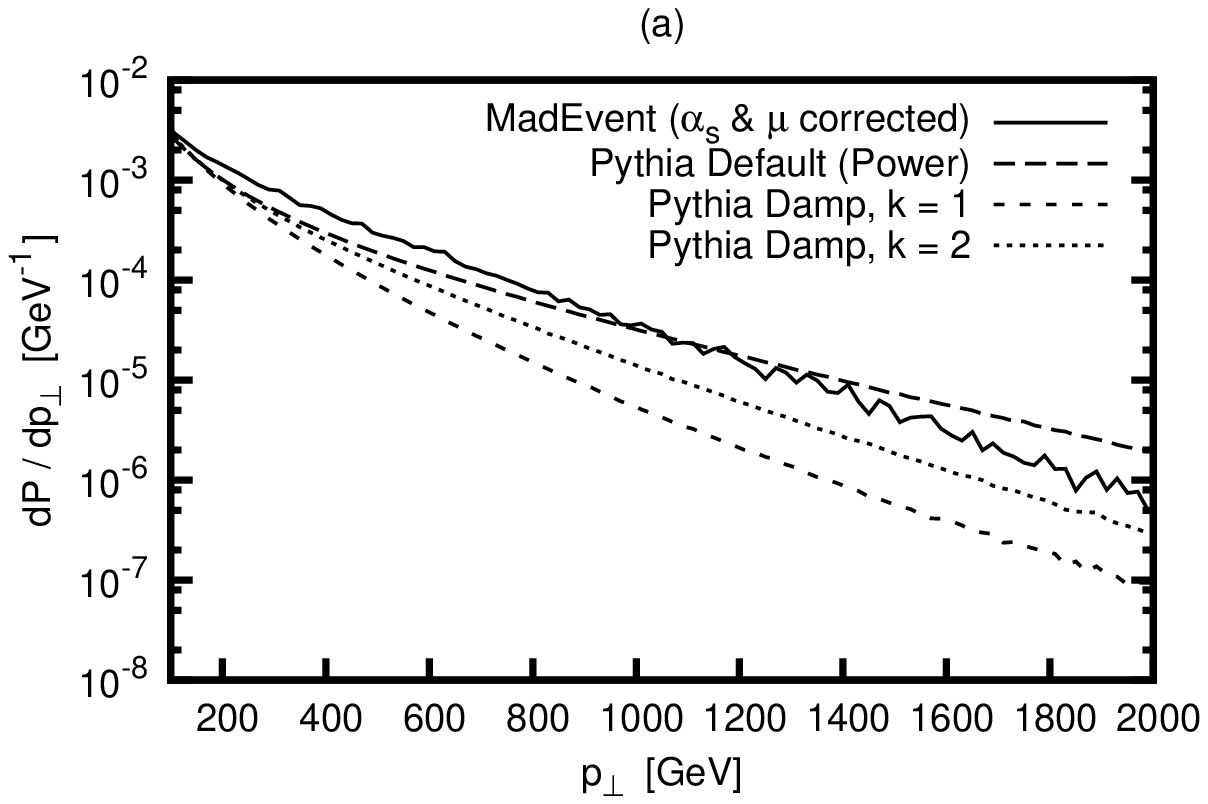}
\includegraphics[angle=270,scale=0.63]{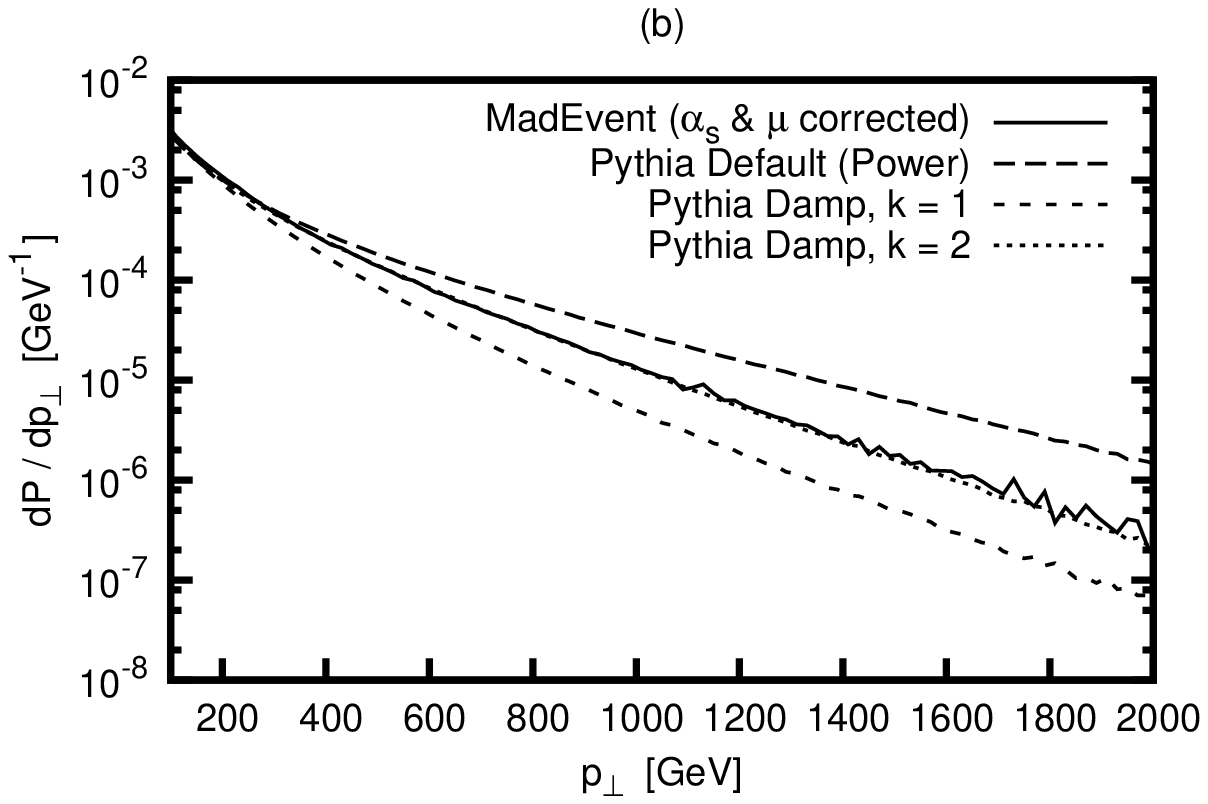}
}
\caption{First emission probability as a function of $\pT$ for (a)
${\squ}_L \, {\schi}_1^0$ and (b) ${\squ}_L \, {\schi}_1^-$
\label{fig:MSSM-pT-3}
}
\end{figure}

\begin{figure}
\centerline{
\includegraphics[angle=270,scale=0.63]{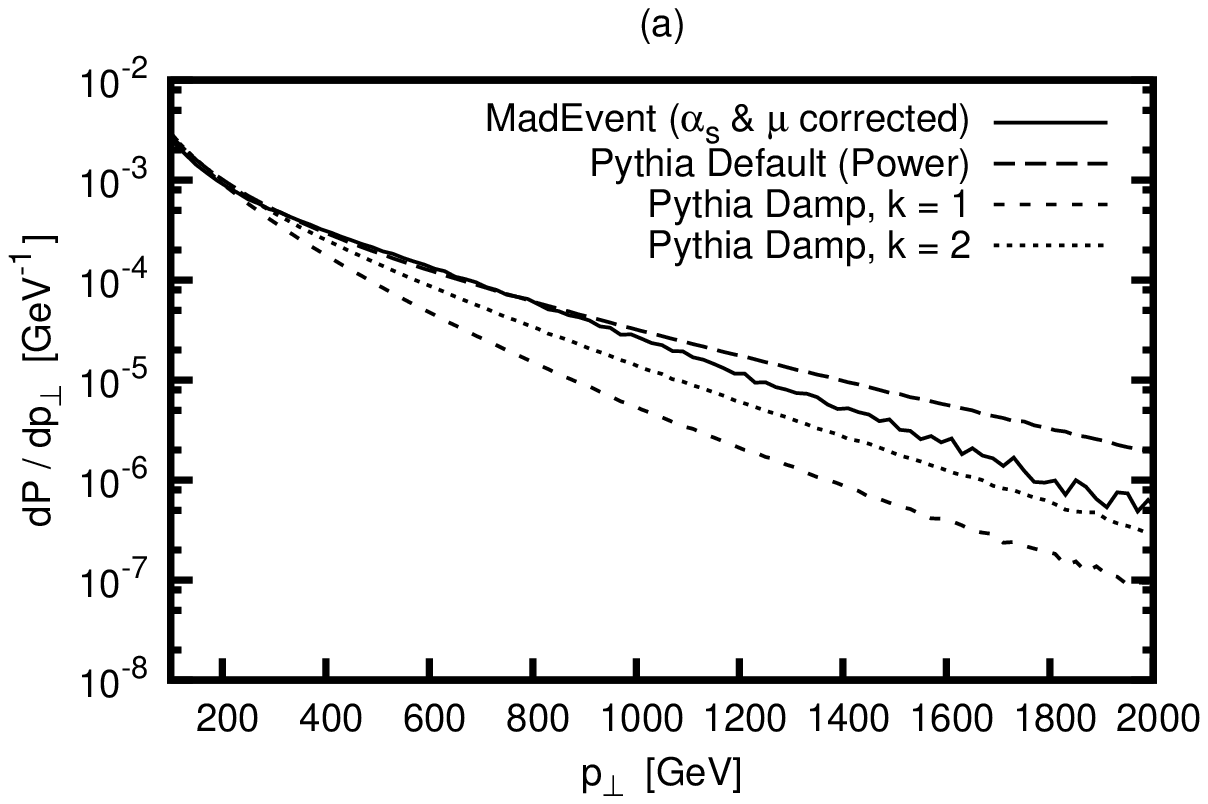}
\includegraphics[angle=270,scale=0.63]{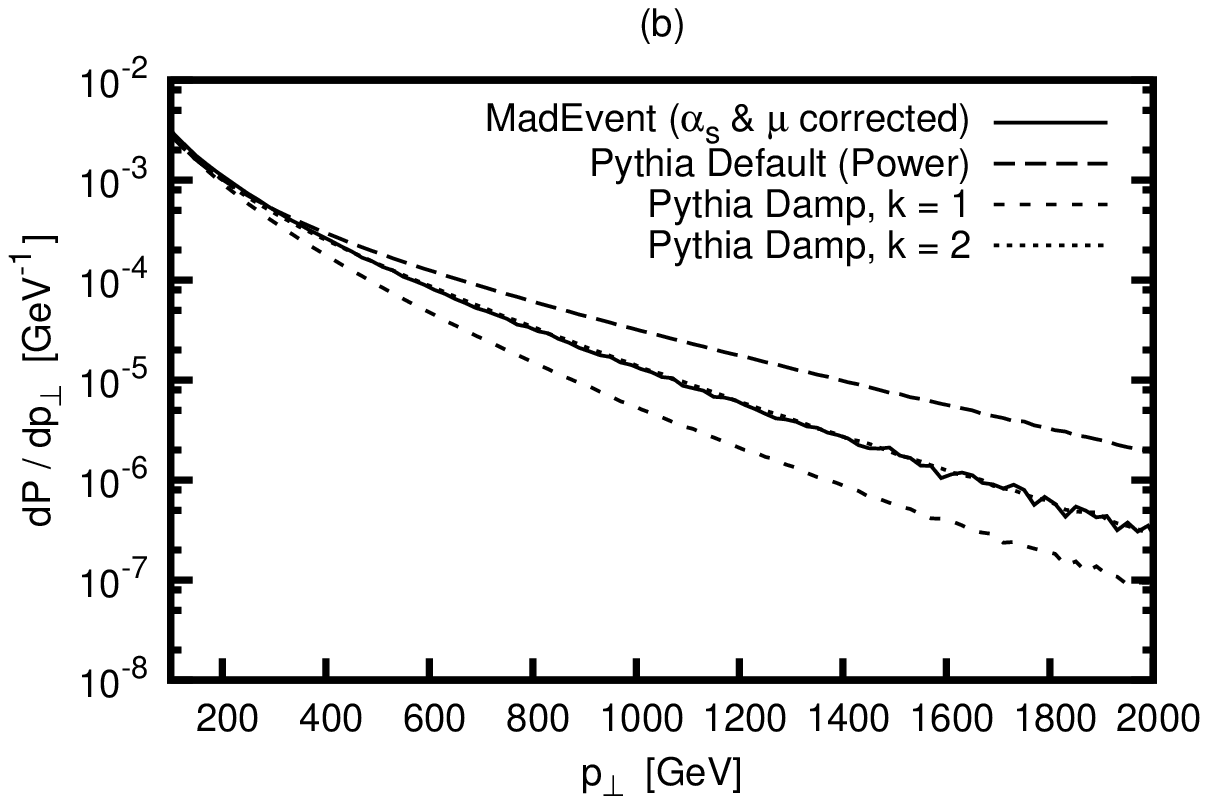}
}
\caption{First emission probability as a function of $\pT$ for 
${\squ}_L \, {\schi}_1^0$, where in (a) resonant graphs have been
removed and (b) the right handed squark masses have been set high
\label{fig:MSSM-pT-4}
}
\end{figure}

Results are shown in Fig.~\ref{fig:MSSM-pT-3} for
(a) ${\squ}_L \, {\schi}_1^0$ and (b) ${\squ}_L \, {\schi}_1^-$.
While the curve for ${\squ}_L \, {\schi}_1^-$ does show the expected behaviour,
with the damping improving the shower description, for
${\squ}_L \, {\schi}_1^0$ we do not obtain the expected results.
Instead, Fig.~\ref{fig:MSSM-pT-4}a shows the results for
${\squ}_L \, {\schi}_1^0$ when resonant graphs are manually set to zero in
the MadEvent generation code. Here, beyond $\pT \sim 1\TeV$,
the tail does appear to fall off with a damped behaviour. Comparing the
real-emission Feynman diagrams for ${\squ}_L \, {\schi}_1^0$ against those
of ${\squ}_L \, {\schi}_1^-$, a key difference is the appearance
of right-handed intermediate squarks. In Fig.~\ref{fig:MSSM-pT-4}b, we
show the results when the right-handed squark masses (${\squ}_R$, ${\sqd}_R$,
${\sqs}_R$ and ${\sqc}_R$) are set high. In this case, we do recover the
behaviour of the ${\squ}_L \, {\schi}_1^-$ result.

\begin{figure}
\centerline{
\includegraphics[angle=270,scale=0.63]{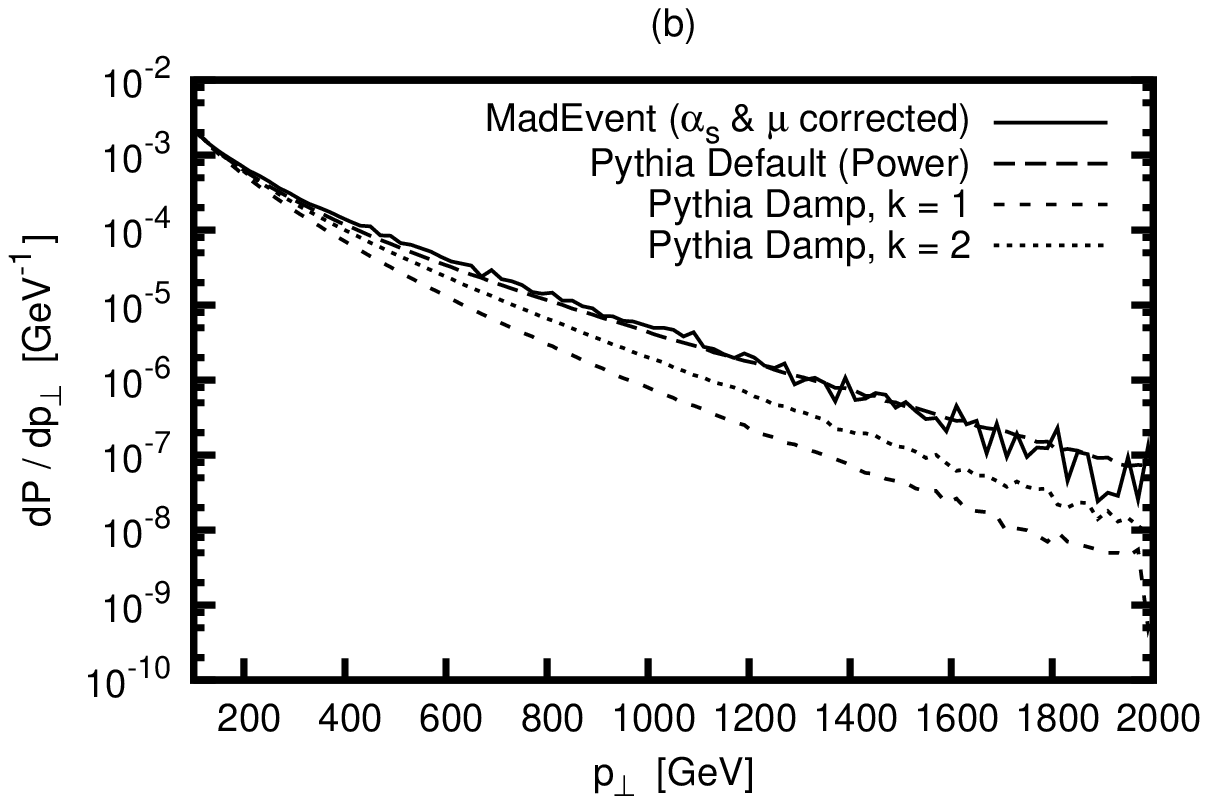}
\includegraphics[angle=270,scale=0.63]{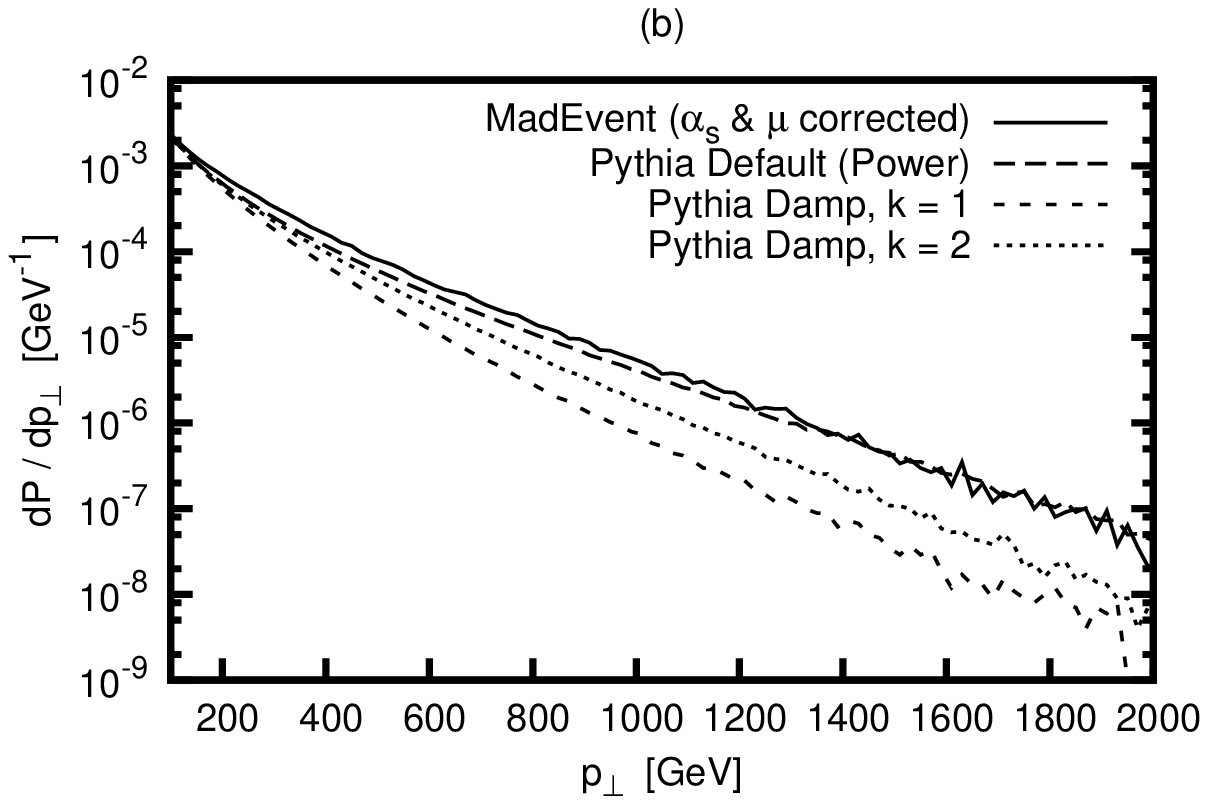}
}
\caption{First emission probability as a function of $\pT$ for
(a) $\sg \, {\schi}_1^0$ and (b) $\sg \, {\schi}_1^-$
\label{fig:MSSM-pT-5}
}
\end{figure}

Finally, in Fig.~\ref{fig:MSSM-pT-5}, we study the processes
(a) $\sg \, {\schi}_1^0$ and (b) $\sg \, {\schi}_1^-$. Again
the expected behaviour is not apparent; both sit closest to the default power
shower curve. 

In summary, for the mixed processes studied in this section,
our ansatz is not an obvious improvement relative to the power shower. 
There is a nontrivial dependence of the emission pattern on the 
SUSY parameter choices that we do not understand, and do not go on 
to study further at this time.

\section{Summary and Outlook}

One direct outcome of this study is an improved matching between
the existing POWHEG program family and the \textsc{Pythia}
$\pT$-ordered showers. The main point is that the \textsc{Pythia}
$\pT$ definition does not quite agree with the normal kinematical
one, although the two do agree closely over most of the phase space
populated by showers. In light of this, the numerical effect of these
improvements are fairly modest, but, in view of the important role that we
foresee for POWHEG-based studies at the LHC, there is every reason to have
the interface to \tsc{Pythia} well understood.

The POWHEG approach, in complete or in simplified form, also allows
us to test the default behaviour of the $\pT$-ordered showers.
In particular we address the issue of ``power'' vs. ``wimpy''
showers, i.e.\ what starting scale to use for the downwards
evolution, specifically for pair production of heavy particles.
Here we show that many processes obey an intermediate behaviour,
where the characteristic $\d\pT^2 / \pT^2$ shower fall-off is
replaced by a steeper $\d\pT^2 / \pT^4$ fall-off for large
$\pT$ values. That is, emissions are allowed up to the kinematical
limit, but at a dampened rate. The damping can be approximated by
a factor
\begin{equation}
\mathcal{P}_{\mrm{dampen}} = \frac{k^2 M^2}{k^2 M^2 + \pT^2} ~,
\end{equation}
for the emission of a parton of transverse momentum $\pT$, where $M^2$
is a characteristic scale of the hard process and $k$ is a fudge factor.

More specifically, we have seen that for coloured pair production, with
$M^2$ set equal to the factorisation scale (noting the slightly different
scale choice used in the \tsc{Pythia} internal processes and the external
MSSM ones) and $k = 2$, we get a reasonable shower behaviour for the
different final states studied here. Instead, when the particles in the
final state of the hard process are colour singlets only, we have seen that
there is no need to impose a damping, consistent with previous results
for single $\W/\Z$ production. The differences between these two cases can
be understood based on the destructive interference between ISR and FSR
that is expected for coloured particles but not for the uncoloured ones.
In no case is the wimpy shower a good choice.

For the mixed coloured/non-coloured final states of Sec.~\ref{sec:mixed},
we have argued that $M^2$ should instead be related to the
coloured partons only, due to this coherence argument. Unfortunately, for
these processes, the results are not as expected. We do not currently
understand the structure of the real-emission matrix elements that gives
rise to the behaviour seen here and, specifically, whether it is an artifact of the
complicated SUSY structure or an inherent property of the radiation pattern.

Not studied here is the case of the production of light coloured
particles, i.e.\ of normal QCD jets. There, the hard process and
the showers produce the same kind of partons, and doublecounting
becomes a main concern. We intend to return to this class of event.

With the advances in computational tools, allowing automatised
(Born-level) higher-order calculations, one may question the need
for more accurate showers. We believe there are two main points
in favour of improving showers, especially when this can be done
with only a modest effort. One is that new physics scenarios are
continuously being proposed, where higher-order calculations may
be overkill for first studies, but nevertheless a realistic
population over all possible event topologies is useful, even if
off by a factor of two or so in some tails. The other is that trial
showers as a means of obtaining Sudakov factors is a crucial
ingredient of CKKW-L matching schemes, so that the quality of
matching is improved if the quality of the shower is also improved.

\subsection*{Acknowledgments}
This work was supported in by the Marie Curie Early Stage Training program
``HEP-EST'' (contract number MEST-CT-2005-019626) and in part by the Marie
Curie research training network ``MCnet'' (contract number
MRTN-CT-2006-035606).

\bibliography{ips}{}
\bibliographystyle{utcaps}

\end{document}